\documentstyle[seceq,amssymb,graphicx]{jfm}

\ifCUPmtlplainloaded
\else
  \font\tenbmi=cmmib10 at 10pt  \skewchar\tenbmi ='177
  \font\sevenbmi=cmmib10 at 7pt \skewchar\sevenbmi ='177
  \font\fivebmi=cmmib10 at 5pt  \skewchar\fivebmi ='177

  \newfam\bmifam
  \textfont\bmifam=\tenbmi
  \scriptfont\bmifam=\sevenbmi
  \scriptscriptfont\bmifam=\fivebmi
  \def\bmi{\fam\bmifam\tenbmi}
\fi

\ifCUPmtlplainloaded
  \font\sfs = mtssi10 at 10.5pt   % sans-serif italic
  \font\sls = mtssbi10 at 10.5pt  % sans-serif bold maths, slanted
  \font\bit = mtmib10 at 10.5pt \skewchar\bit ='177  % bold math italic
\else
  \font\sfs = cmssi10  % sans-serif slanted
  \font\sls = cmssi10  % sans-serif bold maths, slanted
  \font\bit = cmmib10 \skewchar\bit ='177  % bold math italic
\fi

\ifCUPmtlplainloaded
\else
  \font\eurmten=eurm10
  \font\eurmseven=eurm10 at 7pt
  \font\eurmfive=eurm10 at 5pt
  \newfam\eurmfam
  \textfont\eurmfam=\eurmten
  \scriptfont\eurmfam=\eurmseven
  \scriptscriptfont\eurmfam=\eurmfive
  \edef\eurm{\hexnumber\eurmfam}

  \mathchardef\upi="0\eurm19      % for upright Greek character
  \mathchardef\umu="0\eurm16      % for upright Greek character
  \mathchardef\upartial="0\eurm40 % for upright Greek character

  \font\msxten=msam10
  \font\msxseven=msam10 at 7pt
  \font\msxfive=msam10 at 5pt
  \newfam\msxfam
  \textfont\msxfam=\msxten
  \scriptfont\msxfam=\msxseven
  \scriptscriptfont\msxfam=\msxfive
  \edef\msx{\hexnumber\msxfam}

  \mathchardef\leqslant="3\msx36
  \mathchardef\geqslant="3\msx3E

  \let\leq=\leqslant
  \let\geq=\geqslant

\fi

\newcommand{\be}{\begin{equation}}
\newcommand{\ee}{\end{equation}}
\newcommand{\lb}{\label}
\newcommand{\ol}{\overline}
\newcommand{\wt}{\widetilde}

\newcommand{\bk}{{\bmi k}}
\newcommand{\br}{{\bmi r}}
\newcommand{\bn}{{\bmi n}}
\newcommand{\bu}{{\bmi u}}
\newcommand{\bv}{{\bmi v}}
\newcommand{\bx}{{\bmi x}}
\newcommand{\obu}{\overline{\bmi u}}
\newcommand{\bJ}{{\bmi J}}

\newcommand{\bhz}{\widehat{\bmi z}}

\newcommand{\bI}{\mbox{\sls I}}

\newcommand{\bD}{\mbox{\sls D}}
\newcommand{\bR}{\mbox{\sls R}}
\newcommand{\bS}{\mbox{\sls S}}
\newcommand{\bT}{\mbox{\sls T}}

\newcommand{\sD}{\mbox{\sfs D}}
\newcommand{\sS}{\mbox{\sfs S}}
\newcommand{\sT}{\mbox{\sfs T}}

\newcommand{\lap}{\bigtriangleup}
\newcommand{\bdot}{{\mbox{\boldmath $\cdot$}}}
\newcommand{\grad}{{\mbox{\boldmath $\nabla$}}}
\newcommand{\bzed}{{\mbox{\boldmath ${\rm o}$}}}
\newcommand{\bomega}{{\mbox{\boldmath $\omega$}}}

\newcommand{\brho}{{\mbox{\boldmath $\rho$}}}
\newcommand{\bvrho}{{\mbox{\boldmath $\varrho$}}}
\newcommand{\bphi}{{\mbox{\boldmath $\phi$}}}
\newcommand{\bpsi}{{\mbox{\boldmath $\psi$}}}
\newcommand{\btau}{{\mbox{\boldmath $\tau$}}}
\newcommand{\bxi}{{\mbox{\boldmath $\xi$}}}
\newcommand{\bdots}{{\mbox{\boldmath $:$}}}
\newcommand{\btimes}{{\mbox{\boldmath $\times$}}}

\newcommand{\varPi}{\mathit{\Pi}}
\newcommand{\varGamma}{\mathit{\Gamma}}
\newcommand{\varLambda}{\mathit{\Lambda}}
\newcommand{\varXi}{\mathit{\Xi}}
\newcommand{\varOmega}{\mathit{\Omega}}

\newcommand{\varbXi}{{\mbox{\boldmath $\varXi$}}}
\newcommand{\varbOmega}{{\mbox{\boldmath $\varOmega$}}}

\newtheorem{Ex}{Example}

\long\def\symbolfootnote[#1]#2{\begingroup%
\def\thefootnote{\fnsymbol{footnote}}\footnote[#1]{#2}\endgroup}

\begin{document}
\title[Multi-scale Gradient Expansion]
{Multi-Scale Gradient Expansion of the Turbulent Stress Tensor}
\author[Gregory L. Eyink]
{G\ls R\ls E\ls G\ls O\ls R\ls Y\ns L.\ns E\ls Y\ls I\ls N\ls K}
\affiliation{Department of Applied Mathematics \& Statistics,
The Johns Hopkins University, Baltimore, MD 21218}

\date{ }
\maketitle
\begin{abstract}
Turbulent stress is the fundamental quantity in the filtered equation
for large-scale velocity that reflects its interactions with small-scale
velocity modes. We develop an expansion of the turbulent stress tensor
into a double series of contributions from different scales of motion and
different orders of space-derivatives of velocity, a Multi-Scale Gradient
(MSG) expansion. We compare our method with a somewhat similar expansion
---due to Yeo and Bedford, and to Leonard---that is based instead on
defiltering. Our MSG expansion is proved to converge to the exact stress,
as a consequence of the locality of cascade both in scale and in space.
Simple estimates show, however, that the convergence rate may be slow for
the expansion in spatial gradients of very small scales. Therefore, we
develop an approximate expansion, based upon an assumption that similar
or `coherent' contributions to turbulent stress are obtained from disjoint
subgrid regions. This Coherent-Subregions Approximation (CSA) yields
an MSG expansion that can be proved to converge rapidly at all scales and
is hopefully still reasonably accurate. As an important first application
of our methods, we consider the cascades of energy and helicity in
three-dimensional turbulence. To first order in velocity-gradients,
the stress has three contributions: a tensile stress along principal
directions of strain, a contractile stress along vortex lines, and
a shear stress proportional to `skew-strain.'  While vortex-stretching
plays the major role in energy cascade, there is a second, less
scale-local contribution from `skew-strain'. For helicity cascade
the situation is reversed, and it arises scale-locally from `skew-strain'
while the stress along vortex-lines gives a secondary, less scale-local
contribution. These conclusions are illustrated with simple exact
solutions of 3D Euler equations. In the first, energy cascade occurs
by Taylor's mechanism of stretching and spin-up of small-scale vortices
due to large-scale strain. In the second, helicity cascade occurs by
`twisting' of small-scale vortex filaments due to a large-scale screw.
\end{abstract}

\section{Introduction}

It is well-recognized that turbulent cascades are essentially multi-scale
phenomena, and involve the coupling of modes at distinct scales.
The property of {\it locality} implies that these interactions are
mainly between adjacent scales (\cite{Eyink05}). Of course, this locality
is rather weak and modes at scales differing by an order of
magnitude from a fixed scale can make a substantial contribution
to transfer across that scale. In the filtering approach (\cite{Germano92})
the nonlinear interaction between scales is embodied in the {\it turbulent
stress tensor} $\btau$ that appears in the equation for the velocity
$\obu,$ low-pass filtered at length-scale $\ell.$ This stress is the
contribution
to spatial transport of large-scale momentum generated by quadratic
self-coupling
of the subfilter scales. Because of the locality property, the subfilter modes
that
contribute most to the stress are those at scales only somewhat smaller than
$\ell.$
This fact raises the hope that a {\it constitutive relation} may be constructed
for the stress, if the adjacent small scales can be somehow estimated
from the resolved modes. Unfortunately, it is not hard to see that eliminating
the small scales produces a stress whose dependence upon the large-scale
velocity $\obu$ is, in general, spatially nonlocal, history-dependent, and
stochastic (\cite{Lindenberget87,Eyink96b}). Thus, an exact constitutive
relation
for the turbulent stress is formally available but it is quite unwieldy and
not of direct practical utility.

In apparent contradiction to these remarks, several authors have developed
a closed constitutive formula for the turbulent stress
(\cite{YeoBedford88,Leonard97},
Carati, Winckelmans \& Jeanmart (2001)). The basic idea of their approach is to
{\it defilter} $\obu$ to obtain the unfiltered velocity field $\bu$ and then to
use
the latter to calculate the stress. For many common filter kernels, it is
possible
to evaluate the resulting formula as a concrete expansion in powers of the
filtered
velocity-gradients.  Although nothing is proved about the convergence of this
series,
it seems in principle to provide a solution to the `closure problem' of
turbulence.
However, as we argue at length below, this solution is illusory for several
reasons.
Most obviously, defiltering is not defined for all filter kernels. Furthermore,
the defiltering operator, even when defined, is unbounded on the natural
function spaces for the velocity field (e.g. finite-energy functions). Thus,
the convergence cannot hold in general.

Nevertheless, it is an extremely attractive idea to develop an expansion for
the stress in powers of the filtered velocity-gradients. Similar expansions
have proved useful in many areas of physics, e.g. for one-particle
distribution functions in the solution of the Boltzmann equation
(\cite{Enskog17,
ChapmanCowling39}) or for Ginzburg-Landau free energies of superconductors
(\cite{Gorkov59,Tewordt65}). We shall here develop a convergent gradient
expansion for the turbulent stress. It is somewhat more intricate than the
expansion developed in \cite{YeoBedford88,Leonard97} and \cite{Caratietal01},
since
it is expressed by a summation simultaneously over the order of space gradients
and over an integer index indicating the scale of motion involved. Thus, it is
a {\it multi-scale gradient expansion}. This series expansion will be proved
below
to converge, as a consequence of the locality of the turbulent cascade both
in space and in scale. Of course, the rate of convergence may be slow,
especially
for the Taylor expansion in space of small-scales, so that very high order
gradients
could be required to obtain an accurate result. We diagnose the reasons for
potentially poor convergence, and, on that basis, develop also an approximate
expansion which will converge rapidly at all scales. This approximation
may give reasonable accuracy with just a few low-order gradients.

Because it is multi-scale, the expansion considered here does not by itself
give
a closed `constitutive' relation for the stress. However, it may be a useful
point of departure in developing a closure for the stress, if supplemented
with a scheme to estimate subfilter velocity-gradients in terms of filtered
velocity-gradients. This in line with some large-eddy simulation (LES)
approaches
which construct subgrid-stress models by creating `surrogate' subgrid modes.
See \cite{DomaradzkiSaiki97,MisraPullin97,ScottiMeneveau99,Burtonetal02},
and, for an extensive review, \cite{DomaradzkiAdams02}. Our emphasis in this
paper is on fundamental physics rather than on closure models, but we hope
to pursue this in future work.
% \cite{Ey06}.???????????????
Even without a closure prescription, the formula we develop for the stress
makes
many testable predictions. Concrete conclusions will be deduced here for the
joint
cascade of energy and helicity in three space dimensions (3D) and, in a
following
paper [\cite{Ey05b}], for the inverse energy cascade predicted in two
dimensions
by \cite{Kraichnan67}.

The contents of this paper are as follows: In Section 2 we briefly review
the filtering approach to turbulence. In the main Section 3 we develop our
Multi-Scale Gradient (MSG) expansion for the turbulent stress. In Section 4
we develop a more rapidly convergent but less systematic approximation, which
we call the Coherent-Subregions Approximate Multi-Scale Gradient (CSA-MSG)
expansion.
In Section 5 we present the application of our method to 3D energy and helicity
cascades. Technical proofs and calculations are given finally in four
Appendices.

% \newpage

\section{Filtering Approach and Turbulent Stress}

We first give a general discussion of the mechanics of energy transfer
between scales in a turbulent flow. Following \cite{Germano92},
we resolve turbulent fields simultaneously in space and in scale using
a simple filtering approach. We consider initially an
arbitrary dimension $d$ of space. Thus, we define a low-pass filtered
velocity
\be \ol{\bu}(\bx)  = \int d^d\br \,\,G_\ell(\br) \bu(\bx+\br),
    \label{LPfilter} \ee
where $G$ is a smooth mollifier or filtering function, nonnegative, spatially
well-localized, with unit integral $\int d^d\br \,\,G(\br)=1. $ The function
$G_\ell$ is rescaled with $\ell,$ as $G_\ell(\br) = \ell^{-d}G(\br/\ell).$
Likewise, we can define a complementary high-pass filter by
\be \bu^{\prime}(\bx) = \bu(\bx)-\ol{\bu}(\bx).  \label{HPfilter} \ee
If the above filtering operation is applied to the incompressible
Navier-Stokes equation
\be \partial_t\bu + (\bu\bdot\grad)\bu = -\grad p + \nu\lap\bu,  \lb{NSeq} \ee
with $\grad\bdot\bu=0$ determining the pressure $p,$ then one obtains
\be \partial_t\ol{\bu} + (\ol{\bu}\bdot\grad)\ol{\bu} = -\grad\bdot\btau
                -\grad \ol{p} + \nu\lap\ol{\bu},  \lb{LPNSeq} \ee
where
\be \btau  = \ol{\bu\bu}-\ol{\bu}\,\ol{\bu} \lb{stress} \ee
is the {\it stress tensor} from the scales $<\ell$ removed by the filtering.

The equation for energy balance in the large scales is
(\cite{PCML91,Eyink95b}):
\be \partial_t e + \grad\bdot\bJ = -\varPi - \nu|\grad \ol{\bu}|^2,
     \lb{LP-energy-eq} \ee
with large-scale {\it energy density} $e=(1/2)\ol{\bu}^2,$
spatial {\it energy transport} vector in the large scales $\bJ =
(e+\ol{p})\ol{\bu} + \ol{\bu}\bdot\btau -\nu \grad e,$ and scale-to-scale
{\it energy flux}
\be \varPi = -\grad\ol{\bu}\,\bdots\btau. \lb{energy-flux} \ee
The latter quantity is the rate of work done by the large-scale
velocity gradient against the small-scale stress. Of course, it may
be rewritten in various equivalent forms as
\be \varPi = -\grad\ol{\bu}\,\bdots\btau^{{\,\!}^{\!\!\!\!\circ}}
        = -\ol{\bS}\,\bdots\btau
        = -\ol{\bS}\,\bdots\btau^{{\,\!}^{\!\!\!\!\circ}}. \lb{energy-flux-II}
\ee
The first follows from incompressibility of the velocity field, where
\be \btau^{{\,\!}^{\!\!\!\!\circ}} =\btau- ({\rm Tr}\,\btau)\bI/d
\lb{dev-stress} \ee
is the so-called {\it deviatoric stress} (with $\bI$ the $d\times d$ identity
matrix). The second follows from symmetry of the stress tensor, where
\be \ol{\sS}_{ij}=\frac{1}{2}\left(\frac{\partial\ol{u}_i}{\partial x_j}+
                                \frac{\partial\ol{u}_j}{\partial x_i}\right)
\lb{strain}
\ee
is the large-scale strain rate.  The third follows from both properties
combined.

The following formula
\be \btau = \int d^d\br \,G_\ell(\br) \delta \bu(\br)\delta \bu(\br)
              - \int d^d\br \,G_\ell(\br) \delta \bu(\br) \cdot
                \int d^d\br \,G_\ell(\br) \delta \bu(\br). \lb{stress-CET} \ee
represents the stress as a tensor product of velocity increments
$\delta\bu(\br;\bx)=
\bu(\bx+\br)-\bu(\bx)$ averaged over the separation vector $\br$ with respect
to filter function $G_\ell(\br)$ at length-scale $\ell.$ It is easily
verified by multiplying out the increments and integrating
(\cite{Constantinetal94,Eyink95b}). This expression implies, as a direct
consequence,
the matrix positivity of the stress (\cite{Vremanetal94}). It was also the
crucial
point of departure in our discussion of scale locality properties in
\cite{Eyink05}.
This same formula shall play a central role in our development of the
multi-scale
gradient expansion in this work. A decomposition of (\ref{stress-CET}) that we
shall
find useful is
\be   \btau = \bvrho -\bu^{\prime}\bu^{\prime} \lb{stress-CET-decomp} \ee
where
\be \bvrho(\bx) = \int d^d\br \,G_\ell(\br) \delta \bu(\br;\bx)\delta
\bu(\br;\bx),
                                                    \lb{stress-CET-I} \ee
and
\be \bu^\prime(\bx) = - \int d^d\br \,G_\ell(\br) \delta \bu(\br;\bx).
                                                    \lb{stress-CET-II} \ee
It is easy to check that $\bu^\prime(\bx)$ in (\ref{stress-CET-II}) coincides
with the high-pass filtered field in (\ref{HPfilter}), so that
$-\bu^{\prime}\bu^{\prime}$ represents a `fluctuation' contribution to the
subscale stress, while $\bvrho$ represents a `systematic' contribution
from the spatially-averaged, positive-definite, tensor product of
velocity-increments.

\section{Convergent Expansion in Scale and Space}

In this section, we shall develop our convergent expansion for the turbulent
stress. The key to this convergence is the locality of the stress both in scale
and in space. Therefore, we shall discuss in turn these two properties, develop
from them the resulting expansions, and establish their convergence properties.

% \subsection{The First-Order Model}
\subsection{Locality in Scale}

Scale-locality is the property that only modes from length-scales near
the filter scale $\ell$ contribute predominantly to the stress. Recently,
we have given a rigorous proof of this property, assuming only the
inertial-range scaling laws that are observed in experiment and simulations
(\cite{Eyink05}) and we refer to that work for a more complete discussion.
Here we just recall that locality properties were demonstrated there by
introducing a second `test filter' $\varGamma$ and an additional small
length-scale $\delta<\ell.$ A low-pass filtered velocity at scale $\delta$
was then defined by
\be \bu^{>\delta}(\bx)= \int d^d\br \,\,\varGamma_\delta(\br) \bu(\bx+\br).
   \lb{test-filter} \ee
and likewise a stress contribution $\btau^{>\delta},$ arising only from modes
at length-scales $>\delta$, by
\be \btau^{>\delta} = \overline{\bu^{>\delta}\bu^{>\delta}}
      - \overline{\bu^{>\delta}}\,\overline{\bu^{>\delta}}.
    \lb{delta-stress}  \ee
The property of {\it ultraviolet (UV) locality} of the stress is that
\be \lim_{\delta\rightarrow 0}\btau^{>\delta} = \btau. \lb{UV-locality}
\ee
This limit means that modes at extreme subfilter scales ($\delta\ll \ell$)
make little contribution to the stress at scale $\ell.$ The result
(\ref{UV-locality}) was proved in \cite{Eyink05}, with convergence in
a strong $L^p$-norm sense for any $p\geq 1$, under suitable spatial
regularity assumptions on the velocity field. (In addition,
some very mild moment-conditions must be satisfied by the filter
kernels $G$ and $\varGamma;$ see \cite{Eyink05}.) A sufficient condition is
that the scaling exponent $\zeta_{2p}$ of the (absolute) $(2p)$th-order
moment of the velocity-increment $\delta\bu$ should satisfy the bound
$\zeta_{2p}>0$. There is also a similar property of `infrared (IR)
locality', which requires an oppposite condition $\zeta_{2p}<2p.$
See \cite{Eyink05}. However, we shall not need to make use of IR
locality in the present context.
% We just observe that both UV and IR locality of the stress
% in the  $L^1$-norm hold under a natural condition on the exponent of the
% 2nd-order structure function, defined by $\langle |\delta\bu(\br)|^2\rangle
% \sim r^{\zeta_2};$ namely, that $0<\zeta_{2}<2$

\subsubsection{Multi-Scale Decomposition}

We can now reformulate the UV locality property as a multi-scale
expansion of the stress tensor. First, we consider a corresponding
multi-scale decomposition of the velocity field itself.
Let us chose some parameter $\lambda>1,$ e.g. $\lambda=2$ will be our
standard choice. Then consider a geometric sequence of lengths
$\ell_n=\lambda^{-n}\ell,$ for $n=0,1,2,....$ with $\ell_0=\ell$ and
$\ell_n\searrow 0$  as $n\rightarrow\infty. $ For each of these we
can define the corresponding low-pass filtered velocity field
$\bu^{(n)}= \bu^{>\ell_n}$, or
\be \bu^{(n)}(\bx) = \int d^d\br \,\,\varGamma_{\ell_n}(\br) \bu(\bx+\br).
\lb{LP-nth-bu} \ee
This field includes modes at all length-scales down to $\ell_n.$ We can
also define a contribution to the velocity $\bu^{[n]}$ that arises, roughly
speaking, from the length-scales between $\ell_{n-1}$ and $\ell_n,$ by
\be \bu^{[n]}=\bu^{(n)}-\bu^{(n-1)},\,\,\,\,n\geq 1. \lb{nth-scale-bu} \ee
It is convenient to set $\bu^{[0]}=\bu^{(0)}$. In that case,
$\bu^{(n)}=\sum_{k=0}^n \bu^{[k]}$ for $n\geq 0$ and, taking the limit
as $n\rightarrow \infty,$ we get:
\be \bu =\sum_{n=0}^\infty \bu^{[n]}. \lb{bu-MSexpand} \ee
This is the multi-scale decomposition of the velocity field. It is closely
related to other similar scale decompositions, such as multiresolution
expansions in wavelet bases, Paley-Littlewood decompositions, etc.
\cite{Kraichnan74} used a multi-scale decomposition defined by
banded Fourier series in order to discuss locality properties of the
turbulent cascade. The series (\ref{bu-MSexpand}) will converge in an $L^p$
norm, if, for example, the $p$th-order scaling exponent $\zeta_p$ of the
absolute velocity-increment $|\delta\bu(\br)|$ is positive. In fact,
in that case the series (\ref{bu-MSexpand}) has at least a geometric
rate of convergence.

The scale locality proved in \cite{Eyink05} can be restated in the
present terms by defining the stress $\btau^{(n)}=\btau^{>\ell_n},$ or
\be \btau^{(n)}= \overline{\bu^{(n)}\bu^{(n)}}
                                -\overline{\bu^{(n)}}\,\overline{\bu^{(n)}},
\lb{LP-nth-stress} \ee
which includes the contributions from all length-scales $>\ell_n.$ If we
substitute the expansion $\bu^{(n)}=\sum_{k=0}^n \bu^{[k]}$ and take
the limit $n\rightarrow\infty,$ then we obtain a doubly-infinite series
\be \btau = \sum_{n=0}^\infty\sum_{n'=0}^\infty \btau^{[n,n']}
                                       \lb{stress-MSexpand} \ee
with
$ \btau^{[n,n']} = \overline{\bu^{[n]}\bu^{[n']}}
                     -\overline{\bu^{[n]}}\,\,\overline{\bu^{[n']}}. $
% \lb{nn'th-stress}  \ee
This is the desired {\it multi-scale expansion of the stress tensor.} The
term $\btau^{[n,n']}$ represents a stress contribution from one velocity mode
at length-scale $\ell_n$ and another at length-scale $\ell_{n'}.$ The series
(\ref{stress-MSexpand}) converges absolutely in the $L^p$-norm (and, in fact,
at a geometric rate) under the same condition mentioned in the UV locality
statement, namely, the positivity of the scaling exponent of order $2p$,
$\zeta_{2p}>0.$  This is a direct consequence of the concrete estimates
in \cite{Eyink05}.

A remark should be made concerning the limit $n\rightarrow\infty$ in the
expansions (\ref{bu-MSexpand}) and (\ref{stress-MSexpand}) above. For a
finite (but arbitrarily large) Reynolds number, this limit need not be taken,
practically speaking, because the stress contribution from the scales below
the dissipative microscale is negligibly small. Instead, the series can be
truncated at some $n=n_d$ corresponding to the length-scale of the viscous
cutoff. However, for precisely this reason, there is also no difficulty in
taking the limit $n\rightarrow\infty$ at a finite Reynolds number. In fact,
the convergence rate of the expansion for $n$ in the dissipation-range
of scales is greater than in the inertial-range. As shown in \cite{Eyink05}
the UV-locality of the stress depends upon the condition that the H\"older
exponent of the velocity satisfy $\alpha>0,$ and, the larger $\alpha$ may be,
the better this property holds. Since the velocity field is smooth in the
dissipation range ($\alpha=1$), the UV-locality property is correspondingly
stronger there than in the inertial-range where $\alpha<1.$

\subsubsection{Leading Terms and Strong UV-Locality}

Truncation of the series (\ref{stress-MSexpand}) at its leading term
corresponds
to making a {\it strong UV-locality assumption}. In that case, the
exact stress is approximated as
\be \btau\approx \btau^{[0,0]} = \overline{\bu^{[0]}\bu^{[0]}}
                     -\overline{\bu^{[0]}}\,\,\overline{\bu^{[0]}}.
\lb{strong-UVloc} \ee
This approximation neglects all interactions with modes at sub-filter
lengths $<\ell_0=\ell.$ However, all interactions are retained between
the modes at length-scales above the filter scale (both scale-local and
IR scale-nonlocal ones). This approximation achieves closure for the stress
in terms of $\bu^{[0]},$ essentially what is called the Similarity Model
or Bardina  Model in the LES literature (\cite{MeneveauKatz00}). To simplify
our discussion of this leading-order approximation, we may use a special
notation for the low-pass filter of the velocity at scale $\ell$ with
respect to the test kernel $\varGamma$, namely, $\wt{\bu}=\varGamma_\ell*\bu.$
Thus, $\wt{\bu}$ and $\bu^{[0]}$ are different notations for the same
quantity. We introduce also a simplified notation for the complementary
high-pass filter, $\bu''=\bu-\wt{\bu}.$

The approximation (\ref{strong-UVloc}) is rather extreme and the results
in \cite{Eyink05} show that sub-filter scale modes can give a non-negligible
contribution to the stress. Without repeating all the details from that
work, we would like to consider here briefly the magnitude of the error
in making the strong UV-locality assumption. This approximation corresponds
to replacing the exact velocity increment $\delta\bu(\br)$ in the formula
(\ref{stress-CET}) for the stress by
$\delta\bu^{(0)}(\br)=\delta\wt{\bu}(\br)$.
Let us assume that the velocity field has a H\"{o}lder exponent $\alpha,$
that is, $\delta\bu(\br)=O(r^\alpha)$ with $0<\alpha<1.$ (This notation
means, as usual, that there exists a constant $A$ so that $|\delta\bu(\br)|
\leq A r^\alpha,$ or, in a more dimensionally correct form, $|\delta\bu(\br)|
\leq C U (r/L)^\alpha,$ with $L$ the integral length, $U$ the rms velocity,
and $C$ a dimensionless constant.) Then it is not hard to prove that
$\bu''(\bx)=O(\ell^\alpha)$; see \cite{Eyink05} for details. An error
of this magnitude is made for all $\br$ in replacing $\delta\bu(\br)$ by
$\delta\wt{\bu}(\br).$ However, $\delta\bu(\br)=\delta\wt{\bu}(\br)+
\delta\bu''(\br)=\delta\wt{\bu}(\br)[1+O((\ell/r)^\alpha)] \approx
\delta\wt{\bu}(\br)$ for $r\gg\ell.$ Therefore, the substitution of
$\delta\wt{\bu}(\br)$ for $\delta\bu(\br)$ will be relatively accurate
for increments over large separations $r\gg \ell.$ Of course, this
substitution will not be accurate in the opposite case of small separations,
because $\wt{\bu}$ is smooth and thus $\delta\wt{\bu}(\br)\approx O(r)
\ll \delta\bu(\br)\approx O(r^\alpha)$ when $r\ll \ell$. Thus, a relatively
large underestimate results when the strong UV-locality assumption is
applied to velocity-increments at small separations.

Similar results hold for higher-order truncations, for example, for
$\btau^{(n)}$ defined by (\ref{LP-nth-stress}) with $n\geq 1$, or
equivalently,
\be \btau\approx \btau^{(n)}
  = \sum_{k=0}^n\sum_{k'=0}^n \btau^{[k,k']}. \lb{LP-nth-stress-2} \ee
This approximation assumes also UV-locality, but more weakly. It corresponds
to replacing the exact velocity increment $\delta\bu(\br)$ in the formula
(\ref{stress-CET}) for the stress by $\delta\bu^{(n)}(\br)$, with an error
$O(\ell_n^\alpha).$ This substitution will be relatively accurate when
$r\gtrsim \ell_n,$ but not for $r\lesssim \ell_n.$ Here let us note that
(\ref{LP-nth-stress-2}) does not yield a closed formula in quite the same
sense as does (\ref{strong-UVloc}), since it involves all of
the components $\bu^{[k]}$ for $k=0,...,n.$ If one assumes that $G=\varGamma,$
then $\bu^{[0]}=\wt{\bu}=\ol{\bu}$. However, even if $G=\varGamma,$ one cannot
in general obtain the higher terms $\bu^{[1]},...,\bu^{[n]}$ uniquely
from knowledge of $\ol{\bu}.$

\subsection{Locality in Space}

The turbulent stress is {\it a priori} non-local in space. The formula
(\ref{stress-CET}) expresses the stress as an average of velocity increments
over separation vectors, which involves points, in principle, arbitrarily
far away from the considered point. Nevertheless, the stress has some
spatial locality properties, by virtue of the UV scale-locality discussed
in the previous section. The latter property allows one to replace $\bu$
by $\bu^{(n)},$ with an error $O(\ell_n^\alpha)$ that becomes arbitrarily
small for large enough $n.$ The origin of the localness in space is then
the smoothness of the filtered velocity field $\bu^{(n)}$, which will be
even analytic if the filter kernel $\varGamma$ has a compactly supported
Fourier transform. This smoothness allows one to represent filtered increments,
like $\delta\bu^{(n)}(\br),$ by a convergent Taylor-expansion in the separation
vector $\br$. The result is a formula that involves local gradients of the
filtered velocities, i.e. velocity-gradients at the point where the stress
is to be evaluated.

\subsubsection{Gradient Expansion}

Here we develop the gradient expansion for the stress, once contributions have
been
omitted from arbitrarily small-scales. Let us first note the corresponding
expansion
of the filtered velocity $\bu^{(n)}$ itself. This field is smooth and thus the
Taylor
polynomial of degree $m$
\be \delta{\bu}^{(n,m)}(\br;\bx)  =
\sum_{p=1}^m \sum_{p_1+\cdots +p_d=p}
\frac{r_{1}^{p_1}\cdots r_{d}^{p_d}}{p_1!\cdots p_d!}
(\partial_{1}^{p_1}\cdots \partial_{d}^{p_d}\bu^{(n)})(\bx)
= \sum_{p=1}^m \frac{1}{p!} (\br\bdot\grad)^p\bu^{(n)}(\bx)
\lb{Taylor-exp} \ee
converges to the increment $\delta{\bu}^{(n)}(\br;\bx)$ as
$m\rightarrow\infty.$
If the Taylor polynomial $\delta{\bu}^{(n,m)}$ is substituted into
(\ref{stress-CET}),
then it yields
\be \btau^{(n,m)} = \int d^d\br \,G_\ell(\br)
                \delta\bu^{(n,m)}(\br)\delta\bu^{(n,m)}(\br)
              - \int d^d\br \,G_\ell(\br) \delta\bu^{(n,m)}(\br)
                \int d^d\br \,G_\ell(\br) \delta\bu^{(n,m)}(\br),
\lb{tau-approx} \ee
which is our basic approximation to the stress.

An explicit expression is simplest to derive if one assumes that the filter
kernel
$G$ is spherically-symmetric, as we shall do hereafter. In that case, averages
over
the directions of the increment vector $\br$ can be evaluated by a standard
formula
for averages of a product of an even number of vector components over the unit
sphere
in $d$ space dimensions. The result, which is easily proved by induction, is
that
\be \int_{S^{d-1}} \varpi(d\bn)
n_{i_1} n_{i_2}\cdots n_{i_{2p-1}} n_{i_{2p}} =
 \frac{1}{d(d+2)\cdots[d+2(p-1)]}\sum_{\{i_1',i_1''\},...,\{i_p',i_p''\}}
  \delta_{i_1',i_1''}\cdots \delta_{i_p',i_p''} .\lb{iso-avrg-2pth} \ee
where summation is over all of the $(2p-1)!!$ pairings
$\{i_1',i_1''\},...,\{i_p',i_p''\}$
of the $2p$ indices $i_1,i_2,...,i_{2p-1},i_{2p}.$ An average of a product of
an odd number
of unit-vector components is equal to zero. Particular cases are for $p=1$:
\be \int_{S^{d-1}} \varpi(d\bn) n_i n_j = (1/d)\delta_{ij}. \lb{iso-avrg-2nd}
\ee
and $p=2:$
\be \int_{S^{d-1}} \varpi(d\bn)
n_i n_j n_k n_l =
\frac{1}{d(d+2)}[\delta_{ij}\delta_{kl}+\delta_{ik}\delta_{jl}
+\delta_{il}\delta_{jk}].\lb{iso-avrg-4th} \ee
We thus obtain a stress approximation for $m=1:$
\be \tau_{ij}^{(n,1)} = \frac{C_2}{d}\ell^2\frac{\partial u_i^{(n)}}{\partial
x_k}
                               \frac{\partial u_j^{(n)}}{\partial x_k}
\lb{NL-model-1st} \ee
where $C_2=\int d^d\br \,G(\br) \,|\br|^2$ is the 2nd-moment of the
spherically-symmetric
filter function $G.$ Likewise, for $m=2:$
\begin{eqnarray}
 \tau_{ij}^{(n,2)} & = & \frac{C_2}{d}\ell^2\frac{\partial u_i^{(n)}}{\partial
x_k}
                               \frac{\partial u_j^{(n)}}{\partial x_k}
 + \frac{C_4}{2d(d+2)}\ell^4 \frac{\partial^2 u_i^{(n)}}{\partial x_k\partial
x_l}
                             \frac{\partial^2 u_j^{(n)}}{\partial x_k\partial
x_l} \cr
   &  &
\,\,\,\,\,\,\,\,\,\,\,\,\,\,\,\,\,\,\,\,\,\,\,\,\,\,\,\,\,\,\,\,\,\,\,\,\,
\,\,\,\,\,\,\,\,\,
 + \frac{d\cdot C_4-(d+2)C_2^2}{4d^2(d+2)}\ell^4 \bigtriangleup u_i^{(n)}
                                                 \bigtriangleup u_j^{(n)},
\lb{NL-stress-2nd}
\end{eqnarray}
where $C_2$ is as before and $C_4=\int d^d\br \,G(\br) \,|\br|^4$ is the
4th-moment
of the spherically-symmetric filter function $G.$ In these expressions we may
further
substitute the multi-scale decomposition $\bu^{(n)}=\sum_{k=0}^n \bu^{[k]}$ to
obtain
expansions such as for $m=1$
\be \tau_{ij}^{(n,1)} = \frac{C_2}{d}\ell^2 \sum_{l=0}^n\sum_{l'=0}^n
                               \frac{\partial u_i^{[l]}}{\partial x_k}
                               \frac{\partial u_j^{[l']}}{\partial x_k}
\lb{NL-model-1st-2nd} \ee
and similarly for $\btau^{(n,m)}$ with $m>1.$ Thus, we obtain a {\it
multi-scale gradient
(MSG) expansion of the stress} simultaneously in scale and in space.

In Appendix A we prove that $\btau^{(n,m)}$ converges to $\btau^{(n)}$ in the
limit
as $m\rightarrow\infty.$ To keep the proof simple, we establish convergence
in the spatial $L^1$-norm, requiring just finite energy for the velocity-field
$\bu.$
We assume also for the filter kernels that $G$ decays faster than exponentially
in space
and that the Fourier transform $\widehat{\varGamma}$ has compact support. These
specific
assumptions can doubtless be modified in various ways, but they simplify the
details
of the proof. From our discussion of scale-locality in the preceding section,
we recall
that $\btau^{(n)}$ also converges to $\btau$ in the  $L^1$-norm as
$n\rightarrow\infty$,
if the scaling exponent of the 2nd-order structure function satisfies
$\zeta_{2}>0.$
Thus, under these various assumptions, $\btau^{(n,m)}$ converges in the
$L^1$-norm
to the exact stress $\btau$ in the double limit taking first
$m\rightarrow\infty$ and
then $n\rightarrow\infty.$

It is rather rare to be able to show that a systematic turbulence approximation
scheme is convergent. For example, \cite{Kraichnan70} has discussed previous
attempts
to construct expansion schemes based upon Reynolds number, where convergence
has proved quite elusive. A case in point is the gradient-expansion for the
subscale stress proposed in \cite{YeoBedford88,Leonard97} and
\cite{Caratietal01},
based upon defiltering.  As we discuss in Appendix B, there are many fluid
velocity
fields with finite energy---even infinitely smooth ones---for which the
expansion
(\ref{T-exp}) does not converge. The problem becomes more severe the more
rapidly
the Fourier transform of the filter kernel, $\widehat{G}(\bk),$ decays at large
$k.$
Our present study was motivated, in part, by the desire to overcome this
difficulty.
As we have shown (Appendix A and \cite{Eyink05}), the multi-scale gradient
expansion that we have elaborated does indeed converge, under realistic
and rather mild conditions on the turbulent velocity field and with very modest
regularity assumptions on the filter kernels. However, a price has been paid
for this achievement. Unlike the expansion in \cite{YeoBedford88,Leonard97}
and \cite{Caratietal01}, our multi-scale gradient expansion is not closed in
terms of the filtered field $\wt{\bu}=\bu^{[0]},$ but involves also the
subscale
fields $\bu^{[1]},\bu^{[2]},...$ Thus, closure of our expansion requires
an algorithm for estimating these unknown fields.

\subsubsection{Leading Terms and Strong Space-Locality}

Truncation of the gradient expansion $\btau^{(n,m)}$ at small $m$ values---e.g.
approximating $\btau^{(n)}\approx \btau^{(n,1)}$ to first-order in gradients,
as in (\ref{NL-model-1st-2nd})---corresponds to making a strong space-locality
assumption. If the expansion is truncated as well at small values of $n$, then
both UV scale-locality and space-locality are assumed in a strong sense. E.g.,
setting $n=0, m=1$ gives
\be \tau_{ij}^{(0,1)} = \frac{C_2}{d}\ell^2
                         \frac{\partial \wt{u}_i}{\partial x_k}
                         \frac{\partial \wt{u}_j}{\partial x_k}, \lb{NL-model}
\ee
which is the standard first-order Nonlinear Model for the stress
(\cite{Leonard74,Leonard97,BorueOrszag98,MeneveauKatz00}). This observation
gives some insight into the physical approximations underlying that model.

We would like to make an estimate of the error involved in truncating
the gradient expansion to a given order $m$ of space-gradients. As in our
discussion of scale-locality, we shall assume that the velocity field is
H\"{o}lder
continuous, so that $\delta\bu(\br)=O(r^\alpha)$ for some $0<\alpha<1.$ Then
it is not hard to show that the $p$th-order term in the Taylor expansion
(\ref{Taylor-exp}) of $\delta{\bu}^{(n,m)}(\br)$ scales as $(\br\bdot\grad)^p
\bu^{(n)}=O(\ell_n^{\alpha}(r/\ell_n)^p)$ for each $p\geq 1.$ For example, see
\cite{Eyink05}. Compared with the exact increment $\delta\bu(\br)$, we see that
each term is an underestimate for $r\lesssim \ell_n$ and an overestimate for
$r\gtrsim \ell_n,$ and the error is greater for larger $p.$ Truncated to a
given
small order $m,$ the Taylor approximation has the correct order of magnitude
only
for $r\approx \ell_n.$ If we sum over all values of $p,$ to infinite order in
$m,$
then we recover the approximation $\delta{\bu}^{(n)}(\br),$ which we have seen
is an
underestimate for $r\lesssim \ell_n$ but relatively accurate for
$r\gtrsim\ell_n.$

Although this gradient expansion converges, there is no small parameter
involved
(except for $r\ll \ell_n$). The series (\ref{Taylor-exp}) converges only
because
of the inverse factorials $1/p!$ that make coefficients of higher-order terms
quite
small. Thus, we can expect that very large values of $m$ will be required to
make
$\delta{\bu}^{(n,m)}(\br)\approx \delta{\bu}^{(n)}(\br)\approx
\delta{\bu}(\br)$
for $r\gg \ell_n.$ Because the filter $G_\ell(\br)$ is assumed to decay very
rapidly,
increments with $r\gg\ell$ give little contribution to the stress and thus
their poor
approximation is not an issue. However, increments for separations
$\ell_n\lesssim
r \lesssim \ell$ will give a significant contribution. In this interval the
effective
expansion parameter, $r/\ell_n,$ takes values in the range from $1$ for
$r=\ell_n$
up to $\ell/\ell_n=\lambda^n$ for $r=\ell.$ For small $n,$ say $n=0$ or $1,$
the expansion parameter is always O(1) in the relevant interval of $r,$
and the series converges quite rapidly. However, as $n$ increases, the rate
of convergence in $m$ degrades rather quickly. A crude estimate of the size of
$m$ required for an accurate approximation is $m\gg \lambda^n,$ in order for
the terms $O(\lambda^{np}/p!)$ to be small for $p\approx m.$ This estimate
is probably too pessimistic, since it ignores cancellations that will occur
in the average over $\br$ (cf. eq.(\ref{iso-avrg-2pth})). However, we can be
sure
that the $m$ required to obtain $\btau^{(n,m)}\approx \btau^{(n)}$ will
increase
with increasing $n.$

\section{The Coherent-Subregions Approximation}
\lb{multiscale}

We have proved that $\lim_{m\rightarrow\infty}\btau^{(n,m)}=\btau^{(n)},$ but
also argued that larger orders of space-gradients $m$ are required to achieve
this limit for increasing $n.$ This only stands to reason, because increasing
the scale
index $n$ corresponds to adding finer small-scale structure to the velocity
field.
As the velocity field becomes rougher, higher-order terms in the spatial Taylor
expansion
become necessary in order to represent velocity-increments accurately across
fixed
separations.
% Unfortunately, this means that quite high orders of gradients $m$
% might be required to make the approximation $\btau^{(n,m)}$ accurate when $n$
%%is large.
On the other hand, even a first-order expansion of increments,
$\delta\bu^{(n)}(\br)
\approx (\br\bdot\grad)\bu^{(n)},$ is correct on order of magnitude for
separations
$r\approx\ell_n.$ Therefore, one should be able to get a reasonably accurate
approximation by low-order gradients, if one uses such expansions only for this
range of separations where they are order-of-magnitude correct. In the present
section,
we shall use this strategy in order to construct an approximate multi-scale
representation
$\btau^{(n,m)}_*$ for the turbulent stress. Although this modified expansion is
no longer
convergent to the exact stress, it may be more practically useful than the
systematic
approximation $\btau^{(n,m)},$ because it should be more accurate for smaller
orders $m.$

Although it turns out not to be the most serviceable approach, a first natural
idea is
to represent increments $\br$ of length $r\gg \ell_n$ as the sum of end-to-end
increments
across separations of length $\ell_n$ and then to Taylor expand each of the
individual
increments. Thus, defining the unit vector $\widehat{\br}=\br/r,$ one could
write
\be \delta{\bu}^{(n)}(\br;\bx) \approx \sum_{k=0}^{K} \delta{\bu}^{(n)}
      (\ell_n\widehat{\br};\bx+k\ell_n\widehat{\br})
    \approx \ell_n \sum_{k=0}^{K} (\widehat{\br}\bdot\grad){\bu}^{(n)}
      (\bx+k\ell_n\widehat{\br}) \lb{endtoend} \ee
where $K$ is the greatest integer less than or equal to $r/\ell_n$. The formula
(\ref{endtoend}) is likely to be fairly accurate, since the expansion parameter
is $O(1)$ for each term in the sum. However, this expression involves
velocity-gradients evaluated at points $\bx+k\ell_n\widehat{\br}$ on spheres
of radius $k\ell_n$ about $\bx$ for $k=0,1,...,\lambda^n$ and spatially local
expressions do not result for integrals over $\br$ when (\ref{endtoend}) is
substituted into formula (\ref{stress-CET}) for the stress. In order to get a
simple,
local expression, we should instead Taylor-expand always about point  $\bx.$
Thus,
this approach does not lead to the desired result. However, at least it shows
that
an accurate representation of the stress is possible entirely in terms of
filtered
velocity-gradients of low-order, although the representation is spatially
non-local.

To obtain a local representation, we use Taylor expansions around $\bx$, but
only for displacements where they are both rapidly convergent and accurate.
Let us decompose the integrals in (\ref{stress-CET}) into contributions
from `shells'
\be  {\mathcal S}_k = \{\br:\,\ell_{k-1}>|\br|>\ell_k\},
     \,\,\,\,\,\, k\geq 1. \lb{shell} \ee
See Figure 1. Let us also define an `outer shell'
\be {\mathcal S}_0 = \{\br:\,|\br|>\ell_0\} \lb{zero-shell} \ee
and `balls'
\be  {\mathcal B}_k = \{\br:\,|\br|<\ell_{k-1}\},
     \,\,\,\,\,\, k\geq 1. \lb{ball} \ee
formed from unions of the `shells' ${\mathcal S}_{k'}$ with $k'\geq k.$
{}From our earlier discussion we expect that, for $\br\in {\mathcal S}_k,$
\be \delta\bu(\br)  \approx \delta\bu^{[k]}(\br) \approx
\delta\bu^{[k],(m)}(\br).
\lb{shell-approx} \ee
for any Taylor polynomial of degree $m,$ since the expansion parameter is here
$r/\ell_k\approx 1.$ Thus, we can obtain a rapidly convergent Taylor series
expansion if we replace $\delta\bu(\br)$ by $\delta\bu^{(k)}(\br)$ for
$\br\in {\mathcal S}_k.$ However, this replacement implies that modes at
length-scales $\leq \ell_k$ are now represented only for the increments with
$\br\in {\mathcal B}_k$. Furthermore, this ball ${\mathcal B}_k$ of radius
$\ell_{k-1}$ occupies only a fraction $\sim\lambda^{-kd}$ of the total volume
of the region ${\mathcal B}_1$ which effectively contributes to the stress.
Therefore, such a replacement omits important subscale contributions to the
stress. To compensate for this, we can use the calculated stress contribution
from the shell ${\mathcal S}_k$ to estimate crudely the missing part, by
multiplying the calculated contribution with an enhancement factor of
$N_k=\lambda^{kd}.$ This factor represents the number of subregions of volume
$\sim\ell_k^d$ inside the ball ${\mathcal B}_1$ of radius $\ell.$  Multiplying
each $k$-scale contribution in the shell ${\mathcal S}_k$ by the factor $N_k$
amounts to the assumption that each of the subregions gives a similar
or `coherent' contribution to the stress. We shall therefore call this
heuristic estimate the {\it Coherent-Subregions Approximation (CSA)}.
Let us proceed to develop it more systematically.

\begin{figure}
% \vspace{150pt}
\begin{center}
\includegraphics[width=200pt,height=200pt]{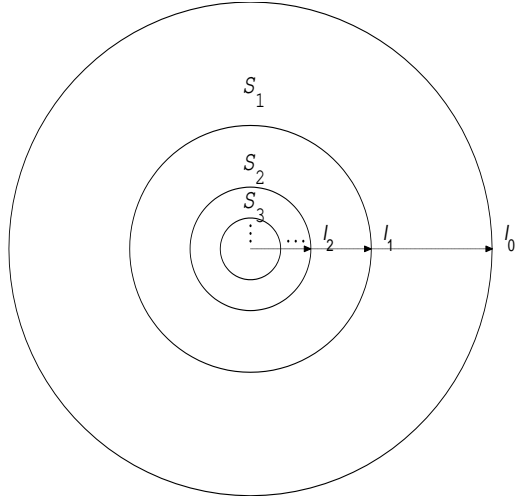}
\end{center}
\caption{{\small {\it Shell Decomposition of Integration Region}. The figure
illustrates the integration region over increment vectors in the stress formula
(\ref{stress-CET}). Most of the contribution comes from the ball of radius
$\ell_0=\ell$ around the point $\bx.$ This ball is decomposed into `shells'
${\mathcal S}_1, {\mathcal S}_2, {\mathcal S}_3,....$ The Taylor expansion of
the $k$-scale contribution to the stress is accurate and rapidly convergent
in the shell ${\mathcal S}_k.$ However, these shells have a smaller portion
of the total volume as $k$ increases.}}

\label{shells}
\end{figure}

It is useful here to employ the formula (\ref{stress-CET-decomp}) which
decomposes the stress into `systematic' part $\bvrho$ and `fluctuation'
part $-\bu^{\prime}\bu^{\prime}.$ These can be represented further by
multiscale decompositions
\be \bvrho = \sum_{k=0}^\infty\sum_{k'=0}^\infty
\bvrho^{[k,k']},\,\,\,\,\,\,\,\,
    \bu^\prime =\sum_{k=0}^\infty \bu^{\prime\,[k]}
                                       \lb{bp-rho-MSexpand} \ee
analogous to (\ref{bu-MSexpand}),(\ref{stress-MSexpand}). Thus,
$\bvrho^{[k,k']}$
represents the contribution from one velocity mode at length-scale $\ell_k$ and
another at length-scale $\ell_{k'}$. The multiscale decomposition of $\bvrho$
can be re-organized as
\be \bvrho = \sum_{k=0}^\infty \bvrho^{[k]} \lb{rho-MSexpand} \ee
where
\be \bvrho^{[k]} = \bvrho^{[k,k]}
+\sum_{l>k}^\infty \left\{\bvrho^{[k,l]}+\bvrho^{[l,k]}\right\}. \lb{rho-kth}
\ee
The term $\bvrho^{[k]}$ represents the contribution arising from a pair of
modes, at least one at length-scale $\ell_k$ and the second at an equal or
smaller length-scale. On the basis of these decompositions we can develop
the desired approximation.

First, let us consider the `systematic' part $\bvrho$. The contributions
$\bvrho^{[k,k']}$ scale as $O(\ell_k^\alpha\ell_{k'}^\alpha)$, when the
velocity
field has H\"{o}lder exponent $\alpha.$ It follows that the dominant term in
$\bvrho^{[k]}$ is the first one on the righthand side of (\ref{rho-kth}),
or $\bvrho^{[k,k]}$. It is possible that the remaining terms sum to a
contribution
of similar order. However, the first term, $\bvrho^{[k,k]},$ is the average
over
space of positive-definite matrices, whereas the remaining terms have no
definite
sign and cancellations can be expected in the summation in (\ref{rho-kth}).
Thus, we expect that
\be \bvrho^{[k]}\approx \bvrho^{[k,k]}
=\int d^d\br\,G_\ell(\br)\delta\bu^{[k]}(\br)\delta\bu^{[k]}(\br).
\lb{missing-1a} \ee
This motivates us to define the CSA value of $\bvrho^{[k]}$, $m$th-order
in gradients, as
\be    \bvrho^{[k],(m)}_* = N_k \int_{{\mathcal S}_k}  d^d\br\,G_\ell(\br)
 \delta\bu^{[k],(m)}(\br)\delta\bu^{[k],(m)}(\br)
\lb{missing-1b} \ee
As discussed earlier, the factor $N_k$  on the righthand side of
(\ref{missing-1b})
corresponds to making the assumption that each subregion of ${\mathcal B}_1$
gives a contribution to the integral (\ref{missing-1a}) similar to that of
the $k$th `shell' ${\mathcal S}_k.$ This is reasonable, since the integrand
is positive-definite and thus there will be little cancellation between the
contributions from the different subregions, which can be expected to add
together
coherently. Using then the replacement (\ref{shell-approx}) in the integral
over
${\mathcal S}_k$ gives (\ref{missing-1b}).  As an additional argument in favor
of the enhancement by $N_k$, let us note that $\bvrho^{[k],(m)}_*$ defined in
(\ref{missing-1b}) with this factor gives an $O(\ell_k^{2\alpha})$ contribution
to the stress, of the correct order of magnitude. See Appendix C.

Similar considerations apply also to the `fluctuation' velocity $\bu^\prime.$
The $k$th-scale contribution $\bu^{\prime\,[k]}$ in (\ref{bp-rho-MSexpand})
is written exactly as
\be \bu^{\prime\,[k]} = -\int d^d\br\,G_\ell(\br)\delta\bu^{[k]}(\br).
\lb{missing-2a} \ee
We then propose the CSA value of $\bu^{\prime\,[k]}$, $m$th-order in
gradients, as
\be \bu^{\prime\,[k],(m)}_*=
 -N_k^{1/2}  \int_{{\mathcal S}_k} d^d\br\,G_\ell(\br)\delta\bu^{[k],(m)}(\br).
\lb{missing-2b} \ee
Note the change in the enhancement factor to $N_k^{1/2}.$ The integrand in
(\ref{missing-2a}) has no definite sign and the $N_k$ different subregions
should
not be expected to contribute coherently. In a work on analytical closures,
\cite{Kraichnan71b} made a similar argument about the shear contribution from
small scales, writing that  `random cancellation effects over the domain $1/k$
in linear dimension should reduce the effective shear of the high wave-numbers
according to the $\sqrt{N}$ law.' Analogous reasoning motivates us to multiply
the righthand side of (\ref{missing-2b}) by $N_k^{1/2}.$

This set of approximations altogether yields the {\it CSA-MSG expansion for the
stress,} $n$th-order in scale index and $m$th-order in gradients:
\be \btau_*^{(n,m)}  = \sum_{k=0}^n \bvrho^{[k],(m)}_*
                        -\sum_{k,k'=0}^n
\bu^{\prime\,[k],(m)}_*\bu^{\prime\,[k'],(m)}_*
\lb{MNL-stress-I} \ee
Using the results for $m=2$ as illustration, we can write
\begin{eqnarray}
\bvrho_{*}^{[k],(2)} & =
& \frac{\ol{C}_2^{[k]}}{d}\ell^2_k\frac{\partial \bu^{[k]}}{\partial x_l}
                               \frac{\partial \bu^{[k]}}{\partial x_l}
+ \frac{\ol{C}_4^{[k]}}{2d(d+2)}\ell^4_k \frac{\partial^2 \bu^{[k]}}{\partial
x_l\partial x_m}
                               \frac{\partial^2 \bu^{[k]}}{\partial x_l\partial
x_m} \cr
     &  & \,\,\,\,\,\,\,\,\,\,\,\,\,\,\,\,\,\,\,\,\,\,\,\,\,\,\,\,\,\,
     + \frac{\ol{C}_4^{[k]}}{4d(d+2)}\ell^4_k \bigtriangleup
\bu^{[k]}\bigtriangleup \bu^{[k]}
\lb{stress-diag}
\end{eqnarray}
and
\be
\bu_{*}^{\prime\,[k],(2)} = \frac{1}{2d\sqrt{N_k}}
                            \ol{C}_2^{[k]}\ell^2_k \bigtriangleup \bu^{[k]}
\lb{stress-offdiag} \ee
As in (\ref{NL-model-1st}),(\ref{NL-stress-2nd}), we have used
(\ref{iso-avrg-2pth})
in order to average over the directions of separation vectors $\br$ in
(\ref{missing-1b}),(\ref{missing-2b}). Note that (\ref{stress-diag}) has the
same
form as (\ref{NL-stress-2nd}) for $\btau^{(k,2)}$ except that the coefficients
are
different. The constants $\ol{C}_p^{[k]}$ for $p=2,4,...$ are the partial
$p$th-moments
of the kernel $G$ over the $k$th `shell' ${\mathcal S}_k$, multiplied by the
factor
$\lambda^{(d+p)k}.$ Expressions are given for these constants in Appendix C,
with $G$ a Gaussian filter.

Our rather rough estimates should obviously be taken with a large grain of salt
and are intended to be accurate qualitatively, but not more than
order-of-magnitude
accurate quantitatively. There is clearly ample room to improve the accuracy of
the
scheme, and many variants and refinements might be fruitfully considered. The
basic
approximation in the `coherent-subregions' assumption, i.e. estimating missing
small-scale contributions to the stress from their effects in subvolumes, will
tend
to enhance the level of fluctuations. However, the CSA stress $\btau^{(n,m)}_*$
in (\ref{MNL-stress-I}) is still likely to be superior to the systematic MSG
expansion
$\btau^{(n,m)}$ when $n\gtrsim 1$ and $m$ is relatively small. The approximate
stress $\btau^{(n,m)}_*$ converges rapidly in the limit $m\rightarrow\infty,$
to some value $\btau^{(n)}_*$ which is, hopefully, a reasonable approximation
of
$\btau^{(n)},$ requiring only moderately large values of $m$ uniformly in $n$.
It achieves our goal of providing a local expression for the stress which
involves
only filtered velocity-gradients of low-order. However, like $\btau^{(n,m)},$
it is not
a proper constitutive relation, because it is not closed in terms of
$\wt{\bu}=\bu^{[0]}.$
Of course, it already makes testable predictions for the stress, if the
smaller-scale
velocity fields $\bu^{[k]}$ for $k\geq 1$ are determined from experiment or DNS
and then
substituted into the model. We report results of such a study elsewhere.
Nevertheless,
an {\it a priori} closure procedure would be useful for modelling purposes.
This could be
accomplished by a stochastic mapping which estimated the velocity gradients
$\grad\bu^{[k]},
\grad\grad\bu^{[k]},$ etc. for $k\geq 1$  from the corresponding gradients for
$k=0.$
We hope to make this the subject of a future work.

\section{The Multi-Scale Gradient Expansion in 3D}

As an application of the general scheme, we shall consider here the turbulent
cascades
of energy and helicity in three space dimensions. In a following work
(\cite{Ey05b}),
we discuss the MSG expansion of the stress for the inverse energy cascade in
two
space dimensions. Many other applications can be considered, such as turbulent
vorticity
transport in the 2D enstrophy cascade (\cite{Ey01,CEEWX03}), or the turbulent
stress
tensor and electromotive force in 3D magnetohydrodynamic cascades. The
technical aspects
of the expansion are similar in all of these cases. As we shall see in this
section,
our method yields a number of interesting predictions for the turbulent stress
in 3D
that may be tested either numerically or experimentally.

\subsection{The Expansion of the Turbulent Stress}

We shall confine ourselves here to considering just the first-order ($m=1$)
term in the gradient-expansion, or $\btau^{(n,1)}$ in (\ref{NL-model-1st}).
We have already noted that this first-order approximation is unlikely to be
very accurate
for larger $n.$ On the other hand, because of the scale-locality of the energy
cascade,
only relatively small values of $n$ need to be considered and thus a
first-order
approximation may be adequate. Furthermore, except for the coefficient,
$\btau^{(n,1)}$
has the same form as the term $\bvrho_{*}^{[n],(m)}$ in (\ref{MNL-stress-I})
for $m=1:$
\be \bvrho_{*}^{[n],(1)} =
     \frac{\ol{C}_2^{[n]}}{d}\ell^2_n\frac{\partial \bu^{[n]}}{\partial x_l}
                               \frac{\partial \bu^{[n]}}{\partial x_l},
\lb{MNL-stress-diag-1st} \ee
and, in addition, the `fluctuation' term in (\ref{MNL-stress-I}) vanishes for
$m=1.$
Thus, the first-order CSA expansion has the closely similar form
\be \btau_{*}^{(n,1)} = \frac{1}{d} \sum_{k=0}^n  \ol{C}_2^{[k]}
                           \ell^2_k\frac{\partial \bu^{[k]}}{\partial x_l}
                                 \frac{\partial \bu^{[k]}}{\partial x_l}.
\lb{MNL-stress-1st} \ee
We expect that $\btau_{*}^{(n,1)}$ in (\ref{MNL-stress-1st}) for large $n$ will
be reasonably accurate in the 3D inertial-range. For example, the estimates
in Appendix C show that the $k$th term in (\ref{MNL-stress-1st}) scales
$\sim O(\ell_k^{2\alpha})$ when the velocity field has H\"{o}lder exponent
$0<\alpha<1.$ This is the correct order of magnitude for the contribution
to the stress from scale $k$ and illustrates the UV locality of the stress.
The series in (\ref{MNL-stress-1st}) then converges at a geometric
rate in the limit $n\rightarrow\infty$ and has the correct overall magnitude
$\sim O(\ell^{2\alpha}).$  By contrast, $\btau^{(n,1)}$ in (\ref{NL-model-1st})
does not have the correct order of magnitude as $n\rightarrow\infty$, but is
too large by a factor of $(\ell/\ell_n)^2.$ This is due to an overestimate
in $\btau^{(n,1)}$ of velocity-increments at large spatial separations, arising
from the first-order Taylor expansion. Thus, our concrete results below
for $\btau^{(n,1)}$ in 3D should be more properly reinterpreted, when $n$ is
large,
for the approximation $\btau_{*}^{(n,1)}$ in (\ref{MNL-stress-1st}) instead.

In any case, we have in 3D the formula for the filtered velocity-gradient
\be \frac{\partial u_i^{(n)}}{\partial x_j}
          = \sS_{ij}^{(n)}-\frac{1}{2}
          \epsilon_{ijk} \omega_k^{(n)} \lb{gradu-S-om}
\ee
in terms of the filtered strain tensor $\bS^{(n)},$ the filtered vorticity
vector
$\bomega^{(n)}$ and the antisymmetric Levi-Civita tensor $\epsilon_{ijk}.$
If (\ref{gradu-S-om}) is substituted into (\ref{NL-model-1st}), it yields
\begin{eqnarray}
\tau_{ij}^{(n,1)} & = & \frac{1}{3}C_2\ell^2\left\{\sS_{ik}^{(n)}\sS_{jk}^{(n)}
+\frac{1}{2}[(\bomega^{(n)}\btimes\bS^{(n)})_{ij}+(\bomega^{(n)}
\btimes\bS^{(n)})_{ji}] \right. \cr
  & &
\,\,\,\,\,\,\,\,\,\,\,\,\,\,\,\,\,\,\,\,\,\,\,\,\,\,\,\,\,\,\,\,\,\,
\,\,\,\,\,\,\,\,\,\,\,\,
+
\left.\frac{1}{4}(\delta_{ij}|\omega^{(n)}|^2-\omega_i^{(n)}
\omega_j^{(n)})\right\}. \lb{stress-NL3D}
\end{eqnarray}
The separate terms in this expression have interesting physical
interpretations.
The first term is proportional to the strain-matrix squared:
\be [\bS^{(n)}]^2 = \sum_{p=1}^3 |\sigma_{p}^{(n)}|^2 {\bmi e}_{p}^{(n)}{\bmi
e}_{p}^{(n)}
    \lb{3d-strain-sqd}, \ee
where $\sigma_p^{(n)}$ and ${\bmi e}_{p}^{(n)}$ are the eigenvalues and
eigenvectors
of the strain matrix $\bS^{(n)}$, satisfying
$\sigma_1^{(n)}+\sigma_2^{(n)}+\sigma_3^{(n)}=0.$
This term represents a tensile stress of magnitude
$(1/3)C_2(\sigma_{p}^{(n)}\ell)^2$
exerted along each of the principal strain directions ${\bmi e}_{p}^{(n)}$ for
$p=1,2,3.$
The last term in (\ref{stress-NL3D}) quadratic in the vorticity likewise
represents
a tensile stress along the two directions orthogonal to the filtered vorticity.
However, the first part of that term proportional to the Kronecker delta
function
is an isotropic stress or turbulent pressure, which does not contribute to the
deviatoric
stress. The other half of the term, proportional to
$\omega_i^{(n)}\omega_j^{(n)},$
is equivalent to a contractile stress of magnitude
$-(1/12)C_2(\omega^{(n)}\ell)^2$
exerted along vortex-lines. Thus, one of the important effects of subscale
modes
is an induced tendency for lines of filtered vorticity $\bomega^{(n)}$ to
resist
lengthening. This `elastic response' of vortex-lines is well-known in other
contexts---for example, turbulence under rapid-distortion (\cite{Crow68}).
However,
the most novel of the stress terms in (\ref{stress-NL3D}) is the middle one,
which
is given by a certain `cross product' of strain and vorticity. More precisely,
we have defined
$(\bomega^{(n)}\btimes\bS^{(n)})_{ij}=\epsilon_{ikl}\bomega_k^{(n)}
\sS_{lj}^{(n)}.$ Note that this tensor is orthogonal to the strain at the same
scale,
$\bS^{(n)}\bdots(\bomega^{(n)}\btimes\bS^{(n)})=0,$ so that we call it the
{\it skew-strain}. The middle term of (\ref{stress-NL3D}), proportional to this
skew-strain, is a sum of shear stresses
\be
\frac{1}{6}C_2 \sum_{p=1}^3 \omega^{(n)}\sigma_p^{(n)}
      \sin\theta_p^{(n)} [\widetilde{{\bmi e}}_{p}^{(n)}
{\bmi e}_{p}^{(n)}+{\bmi e}_{p}^{(n)}\widetilde{{\bmi e}}_{p}^{(n)}]
\lb{stress=shear} \ee
where $\theta_p^{(n)}$ is the angle between $\bomega^{(n)}$ and ${\bmi
e}_{p}^{(n)}$
and $\widetilde{{\bmi e}}_{p}^{(n)}$ is the unit vector orthogonal to both
$\bomega^{(n)}$ and ${\bmi e}_{p}^{(n)},$ given by the righthand rule. If we
introduce the new unit vectors ${\bmi e}_{p\pm}^{(n)}=[{\bmi e}_{p}^{(n)}\pm
\widetilde{{\bmi e}}_{p}^{(n)}]/\sqrt{2},$ then (\ref{stress=shear}) becomes
\be
\frac{1}{6}C_2 \sum_{p=1}^3 \omega^{(n)}\sigma_p^{(n)}
      \sin\theta_p^{(n)} [{\bmi e}_{p+}^{(n)}{\bmi e}_{p+}^{(n)}
                         -{\bmi e}_{p-}^{(n)}{\bmi e}_{p-}^{(n)}]
\lb{stress=shear-45} \ee
Hence, there are both tensile and contractile stresses exerted along the
vectors
${\bmi e}_{p\pm}^{(n)}.$ These are obtained from the strain eigenvector
${\bmi e}_{p}^{(n)}$ by rotating it $\pm\pi/4$ radians around the normal
component of the vorticity vector $\bomega^{(n)}.$

The above vector formalism helps to make clear the geometry of the various
stress contributions. However, it is perhaps more conventional to write
these stresses in terms of the fluid deformation matrix $\bD^{(n)},$
defined by $\sD^{(n)}_{ij}=\partial u^{(n)}_i/\partial x_j,$ and its
symmetric part $\bS^{(n)}$ and anti-symmetric part $\varbOmega^{(n)}.$
Of course, $\bS^{(n)}$ is the strain tensor and $\varbOmega^{(n)}$
is related to the vorticity vector $\bomega^{(n)}$ by the standard relation
$\varOmega^{(n)}_{ij}=-(1/2)\epsilon_{ijk}\omega^{(n)}_k.$ In terms
of the deformation matrix the 1st-order term (\ref{NL-model-1st}) in the MSG
expansion can be written (in fact, in any dimension $d$) as
\be
\btau^{(n,1)} = \frac{1}{d}C_2\ell^2\bD^{(n)}[\bD^{(n)}]^\top.
\lb{stress-NL3D-mat}
\ee
The decomposition analogous to (\ref{stress-NL3D}) is then
\be
\btau^{(n,1)} = \frac{1}{d}C_2\ell^2\left\{\bS^{(n)}\bS^{(n)}
+ [\varbOmega^{(n)},\bS^{(n)}]-\varbOmega^{(n)}\varbOmega^{(n)}
\right\}
\lb{stress-NL3D-mat-dcmp}
\ee
where
$[\varbOmega^{(n)},\bS^{(n)}]=\varbOmega^{(n)}\bS^{(n)}-\bS^{(n)}
\varbOmega^{(n)}$
is the commutator matrix. Thus, in 3D, $\bS^{(n)}\bS^{(n)}$ is the
strain-squared
as in (\ref{3d-strain-sqd}), $-\varbOmega^{(n)}\varbOmega^{(n)}$ gives the
tensile
stress in the plane normal to vortex lines, and $[\varbOmega^{(n)},\bS^{(n)}]$
is the `skew-strain.'

\subsection{Energy Cascade in 3D}

It is interesting to consider the consequences of the stress in formula
(\ref{stress-NL3D})
for the energy cascade. When (\ref{stress-NL3D}) is substituted into equation
(\ref{energy-flux-II}) for the energy flux, one gets the following result
to first-order in gradients:
\be \varPi^{(n,1)}
 =  \frac{1}{3}C_2\ell^2\left\{ -{\rm Tr}\,\left(\ol{\bS}(\bS^{(n)})^2\right)
 + \frac{1}{4} (\bomega^{(n)})^\top\ol{\bS}(\bomega^{(n)})
 + \ol{\bS}\bdots (\bS^{(n)}\btimes\bomega^{(n)})\right\}.
 \lb{flux-3D1st} \ee
The middle term has an obvious physical meaning. It represents the rate of work
done
by the filtered strain $\ol{\bS}$ in order to stretch the lines of vorticity
$\bomega^{(n)}$ against the resisting contractile stress of the subscales. This
remarkable relationship between energy flux and {\it vortex-stretching} was
already
observed by \cite{BorueOrszag98} using the Nonlinear Model, which is a special
case
of our result for $n=0$ and $G=\varGamma$:
\be \varPi^{(0,1)}=  \frac{1}{3}C_2\ell^2\left\{ -{\rm Tr}\,(\ol{\bS}^3) 
+ \frac{1}{4}\ol{\bomega}^\top\ol{\bS}\ol{\bomega}\right\}
\lb{flux-NL3D} \ee
Vortex-stretching was suggested long ago by \cite{Taylor38} as the basic
dissipation
mechanism of 3D turbulence. Note that the first term in
(\ref{flux-NL3D}) proportional to the {\it strain skewness} can also be related
to vortex-stretching, on average, using a relation of \cite{Betchov56}.
His result states that for an incompressible fluid
$ -\langle{\rm Tr}\,(\ol{\bS}^3)\rangle
          = (3/4)\langle \ol{\bomega}^\top\ol{\bS}\ol{\bomega}\rangle,$
% \lb{Betchov-3D} \ee
where $\langle\cdot\rangle$ denotes either ensemble average over a
statistically
homogeneous turbulence or any volume-average where boundary terms from
integration-by-parts can be ignored. According to Betchov's relation, precisely
75\% of the mean energy flux in the Nonlinear Model comes from strain skewness
and 25\% from vortex-stretching. It is interesting that Betchov's relation
can be generalized as follows:
\be
-\langle{\rm Tr}\,\left(\ol{\bS}(\bS^{(n)})^2\right)\rangle =
   \frac{1}{2}\langle (\ol{\bomega})^\top\bS^{(n)}(\bomega^{(n)})\rangle
    + \frac{1}{4} \langle(\bomega^{(n)})^\top\ol{\bS}(\bomega^{(n)})\rangle.
\lb{Betchov-3D-gen} \ee
This result is proved in our Appendix D under the same assumptions as
Betchov's.
% This result can also be rewritten using a generalized Betchov
% relation (Appendix C) as
% \be \varPi^{(n,1)}
% =  \frac{1}{3}C_2\ell^2\left\{
%%\frac{1}{2}(\ol{\bomega})^\top\bS^{(n)}(\bomega^{(n)})
%  + \frac{1}{2} (\bomega^{(n)})^\top\ol{\bS}(\bomega^{(n)})
% + \bomega^{(n)}\bdot (\ol{\bS}\btimes\bS^{(n)})\right\}
% \lb{flux-3D2nd} \ee
By using (\ref{Betchov-3D-gen}) the first term in (\ref{flux-3D1st}) can be
related to vortex-stretching in general for all $n$. However, the last term
in (\ref{flux-3D1st}) appears to be fundamentally different. It appears
only due to contributions of subscales and there is no analogue in the
Nonlinear Model (\ref{flux-NL3D}) for $n=0.$ Some additional insight on that
term can be obtained by rewriting $\ol{\bS}\bdots
(\bS^{(n)}\btimes\bomega^{(n)})
=\bomega^{(n)}\bdot (\ol{\bS}\btimes\bS^{(n)}),$ where
$\ol{\bS}\btimes\bS^{(n)}$
is the dual vector corresponding to the antisymmetric commutator matrix
$[\ol{\bS},\bS^{(n)}],$ i.e.
$(\ol{\bS}\btimes\bS^{(n)})_i=(1/2)\varepsilon_{ijk}
([\ol{\bS},\bS^{(n)}])_{jk}.$ Thus, this new term arises from rotation of the
subscale strain $\bS^{(n)}$ relative to the filtered strain $\ol{\bS},$ and
vanishes if the orthogonal eigenframes of these two symmetric matrices
coincide.

Using the CSA stress in (\ref{MNL-stress-1st}) one gets a similar
result as (\ref{flux-3D1st}):
\be \varPi^{(n,1)}_* = \frac{1}{3} \sum_{k=0}^n  \ol{C}_2^{[k]}
   \ell^2_k\left\{ -{\rm Tr}\,\left(\ol{\bS}(\bS^{[k]})^2\right)
 + \frac{1}{4} (\bomega^{[k]})^\top\ol{\bS}(\bomega^{[k]})
 + \ol{\bS}\bdots (\bS^{[k]}\btimes\bomega^{[k]})\right\}. \lb{flux-MM-3D1st}
\ee
The remarks we have made above on physical interpretation of $\varPi^{(n,1)}$
apply
equally here. However, the sum in (\ref{flux-MM-3D1st}) also has a limit for
large $n$ and we expect that it gives a quite reasonable model for energy flux
in 3D.
 Using the generalized Betchov relation (\ref{Betchov-3D-gen}) the CSA mean
flux
can be written as
\be \langle \varPi^{(n,1)}_*\rangle
   = \frac{1}{3} \sum_{k=0}^n  \ol{C}_2^{[k]}\ell^2_k\left\{
   \frac{1}{2}\langle (\ol{\bomega})^\top\bS^{[k]}(\bomega^{[k]})\rangle
  + \frac{1}{2} \langle(\bomega^{[k]})^\top\ol{\bS}(\bomega^{[k]})\rangle
  + \langle \ol{\bS}\bdots (\bS^{[k]}\btimes\bomega^{[k]})\rangle\right\}.
\lb{mean-flux-MM-3D1st} \ee
The first two terms in each summand arise from vortex-stretching and will tend
to
be positive, certainly for small $k$ when $\bS^{[k]}\propto
\ol{\bS},\bomega^{[k]}
\propto \ol{\bomega}.$ On the other hand, the third term from skew-strain then
nearly
vanishes. For the latter term to be important, there must be some
characteristic rotation
of $\bS^{[k]}$ relative to $\ol{\bS}$ as $k$ increases. Of course, it is not
hard to
see that each term is zero on average when $\bS^{[k]},\bomega^{[k]}$ are
uncorrelated with
$\ol{\bS}$, which must be expected in the limit as $k\rightarrow\infty.$
Therefore,
it only for intermediate values of $k$ that the third term can contribute
to mean energy flux.

The physical mechanism of energy cascade by these stress terms can be
illustrated by the following:

\begin{Ex}
Vortex Tube Stretched by a Constant Strain
\end{Ex}

\begin{figure}
% \vspace{150pt}
\begin{center}
\includegraphics[width=250pt,height=200pt]{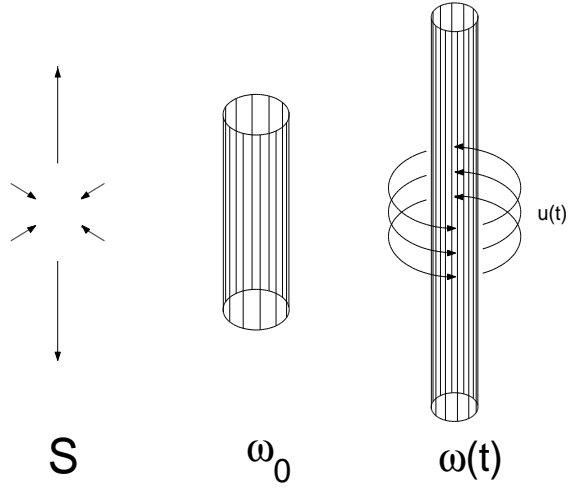}
\end{center}
\caption{{\small {\it Energy Cascade by Vortex-Stretching}. The figure
illustrates
a cylindrical tube of parallel vortex lines (center) in a constant strain field
with stretching direction along the vortex axis (left). The result, shown on
the
right, is that the vortex is stretched and its cross-sectional area shrunken.
The `spin-up' of the vortex increases its energy and generates more positive
(tensile) stress in the plane perpendicular to the vortex axis.}}
\label{stretch}
\end{figure}

We consider a small-scale cylindrical vortex-tube, parallel to the $z$-axis,
with circular cross-section of radius $R=\ell_k$ and with vorticity magnitude
$\omega_0=\omega^{[k]}.$ This is an exact stationary solution of the 3D Euler
equation with two-dimensional symmetry. Thus, it may be described by a
pseudoscalar
stream function
\be \psi^{[k]}(x,y) = \left\{\begin{array}{ll}
                              \,\,\,\,\,(1/4)\omega_0[R^2-r^2] & r<R \cr
                              -(1/2)\omega_0 R^2\ln(r/R) & r>R
                             \end{array} \right.  \lb{vortex-tube}  \ee
where $r$ is the radial distance from the $z$-axis in cylindrical coordinates
$(r,\theta,z).$ The corresponding velocity field is
\be \bu^{[k]}(x,y) = \left\{\begin{array}{ll}
                               (1/2)\omega_0\bhz \btimes \br & r<R \cr
                               (1/2)\omega_0(R/r)^2 \bhz \btimes \br & r>R
                             \end{array} \right.  \lb{vortex-tube-vel}  \ee
where $\bhz$ is the unit vector in the $z$-direction. See Figure 2. This
small-scale field is now superimposed with a large-scale velocity
\be \ol{\bu}=\left[\begin{array}{c}
                   -(\ol{\sigma}/2)x \cr
                   -(\ol{\sigma}/2)y \cr
                     \ol{\sigma}z
                    \end{array} \right], \ee
with deformation matrix $\ol{\bD}$
\be \ol{\bD}=\left[\begin{array}{ccc}
                   -\ol{\sigma}/2 & 0 & 0 \cr
                   0 & -\ol{\sigma}/2 & 0 \cr
                   0 & 0 & \ol{\sigma}
                    \end{array} \right]. \ee
This is a pure large-scale strain $\ol{\bD}=\ol{\bS}$ with vorticity
$\ol{\bomega}=\bzed.$ If $\ol{\sigma}>0,$ then this corresponds to
an axisymmetric stretching along the $z$-direction and compression in
the other two directions (Figure 2). The combination of the large-scale
and small-scale fields gives an exact solution of the 3D Euler equation
$(\partial_t+\bu\bdot\grad)\bomega^{[k]}(t)=(\bomega^{[k]}(t)\bdot\grad)\bu
=0,$ where $\bu=\ol{\bu}+\bu^{[k]}(t)$ and where $\bomega^{[k]}(t)$ is the
same as the initial vorticity field, made time-dependent by the substitutions
$\omega(t)=e^{\ol{\sigma}t}\omega_0$ and $R(t)=e^{-\ol{\sigma}t/2}R.$
For example, see \cite{Neu84}, who considers a more general set of
solutions. In the present case, the small-scale vortex is stretched along
its axis and, by incompressibility, its cross-section shrinks. To conserve
the circulation around the tube, the vorticity and the velocity in the
small-scales correspondingly increase. See Figure 2. This `spin-up' by the
large-scale strain results in a transfer of energy to the small-scales.

The process can be understood from our general formulas above. Without
loss of generality, we can focus on the instantaneous transfer at the
initial time $t=0.$ The velocity-gradient tensor in the small-scales
is then
\be \bD^{[k]}(x,y) = \left\{\begin{array}{cc}
                               \frac{1}{2}\omega^{[k]}
                               \left[\begin{array}{ccc}
                                      0 & -1 & 0 \cr
                                      1 &  0 & 0 \cr
                                      0 &  0 & 0
                                     \end{array} \right]& r<\ell_k \cr

\frac{1}{2}\omega^{[k]}\left(\frac{\ell_k}{r}\right)^2
                               \left[\begin{array}{ccc}
                                      \sin(2\theta)  & -\cos(2\theta) & 0 \cr
                                      -\cos(2\theta) & -\sin(2\theta) & 0 \cr
                                      0 &  0 & 0
                                     \end{array} \right]& r>\ell_k
                             \end{array} \right.  \lb{vortex-tube-D}  \ee
This is purely rotational for $r<\ell_k$ and is a pure strain for $r>\ell_k.$
Substituting into (\ref{MNL-stress-diag-1st}) gives the stress
\be   \btau^{[k],(1)}_*     = \frac{1}{12}\ol{C}^{[k]}|\omega^{[k]}\ell_k|^2
                               \left[\begin{array}{ccc}
                                      1 &  0 & 0 \cr
                                      0 &  1 & 0 \cr
                                      0 &  0 & 0
                                     \end{array} \right] \times
                               \left\{\begin{array}{cc}
                                       1 & r<\ell_k \cr
                                      (\ell_k/r)^4 & r>\ell_k
                                     \end{array} \right.  \lb{vortex-tube-rho}
\ee
to first order in gradients. This result represents the
$-\varbOmega^{[k]}\varbOmega^{[k]}$
term for $r<\ell_k$ and the $\bS^{[k]}\bS^{[k]}$ term for $r>\ell_k$. The
`skew-strain'
vanishes identically for this right cylindrical vortex tube. We see that the
net stress
is tensile in the 2D plane perpendicular to the vortex tube, set up by the
velocity
circulating around the vortex axis. Its deviatoric part includes a contractile
stress along the vortex axis. These stresses oppose the axial stretching and
lateral compression by the large-scale strain, and increase in magnitude as
the vortex spins up. The work of the large-scale strain against these resistive
stresses is the basic mechanism of energy transfer to the small-scales.

It is interesting to observe that if $\ol{\sigma}<0$, then the energy flux
corresponding to (\ref{vortex-tube-rho})
\be   \Pi^{[k],(1)}_*     =
      \frac{1}{12}\ol{C}^{[k]}\ol{\sigma}|\omega^{[k]}\ell_k|^2 \times
                               \left\{\begin{array}{cc}
                                       1 & r<\ell_k \cr
                                      (\ell_k/r)^4 & r>\ell_k
                                     \end{array} \right.
\lb{vortex-tube-Eflux}
\ee
is negative and the large-scale strain `spins down' the small-scale
vortex, by the time-reverse of the process considered above. What is crucial
for foward energy transfer is that the vortex should align with a stretching
direction of the large-scale strain rather than with a shrinking direction. We
know from the relation of \cite{Betchov56} that, in an incompressible flow,
mean
vortex-stretching requires that there be typically two positive strain
eigenvalues
and one negative eigenvalue. This tendency has been confirmed for
dissipation-range
velocity gradients by DNS (\cite{Ashurstetal87}) and for inertial-range
(filtered) velocity-gradients by experiment (\cite{Taoetal02,VanderBos02}).
Furthermore, these empirical studies have shown that the vorticity vector tends
to
align with the intermediate, weakly stretching eigendirection of the strain at
the
same scale, rather than with the strongest stretching direction. Some
theoretical
understanding how this occurs can be obtained from simple Lagrangian dynamical
models
of the velocity-gradients
(\cite{Vieillefosse82,Vieillefosse84,Cantwell92,Chertkovetal99}).
Thus, the situation in turbulence is slightly different from that which we
imagined
in our simple example above. However, the mechanism of the energy transfer
process
appears to be essentially the same.

\subsection{Helicity Cascade in 3D}

It is well-known that 3D smooth solutions of the inviscid, incompressible fluid
equations have in addition to the energy a second quadratic invariant,
the helicity (\cite{Mor,Mof}):
\be H(t) = \int d^3\bx \,\,\bu(\bx,t)\bdot\,\bomega(\bx,t).  \ee
When helicity is input at large scales together with energy, then there is in
3D a joint cascade of both invariants to high-wavenumber (\cite{BFLLM,Kr73}).
The flux of helicity can be expressed quite similarly to the flux of energy
in (\ref{energy-flux}), as
\be \varLambda = -2\grad\ol{\bomega}\bdots\,\btau \lb{helicity-flux} \ee
See \cite{CCE03}. Formula (\ref{helicity-flux}) is quite intriguing, since it
implies
that the stress $\btau$ must be correlated simultaneously with both the
velocity-gradient
and the vorticity-gradient in a joint cascade of energy and helicity. Our work
sheds
some light on how this is achieved.

A vorticity-gradient may be decomposed into symmetric and anti-symmetric parts,
as:
\be \frac{\partial \omega_i}{\partial x_j}
          = R_{ij} + \varXi_{ij} = R_{ij}-\frac{1}{2}\epsilon_{ijk} \xi_k
\lb{gradom-R-xi} \ee
where $\bR=(1/2)[(\grad\bomega)+(\grad\bomega)^\top],\,\,\varbXi=(1/2)
[(\grad\bomega)-(\grad\bomega)^\top]$ and $\bxi=\grad\btimes\bomega.$ We may
write the helicity flux also as $\varLambda=-2\ol{\bR}\bdots\,\btau,$ because
of the symmetry of the stress tensor. Therefore, if we substitute the
first-order stress formula (\ref{stress-NL3D}), then we get the expression
for helicity flux analogous to (\ref{flux-3D1st}):
\be \varLambda^{(n,1)}
 =  \frac{2}{3}C_2\ell^2\left\{ -{\rm Tr}\,\left(\ol{\bR}(\bS^{(n)})^2\right)
 + \frac{1}{4} (\bomega^{(n)})^\top\ol{\bR}(\bomega^{(n)})
 +  \ol{\bR}\bdots\,(\bS^{(n)}\btimes\bomega^{(n)})\right\}
 \lb{Hflux-3D1st} \ee
This is the exact expression to first-order in gradients. Of course, we can
also write down a CSA expansion for helicity flux,
\be \varLambda^{(n,1)}_* = \frac{2}{3} \sum_{k=0}^n  \ol{C}_2^{[k]}
   \ell^2_k\left\{ -{\rm Tr}\,\left(\ol{\bR}(\bS^{[k]})^2\right)
 + \frac{1}{4} (\bomega^{[k]})^\top\ol{\bR}(\bomega^{[k]})
 + \ol{\bR}\bdots(\bS^{[k]}\btimes\bomega^{[k]})\right\},
\lb{Hflux-MM-3D1st} \ee
analogous to (\ref{flux-MM-3D1st}) for energy flux. All the stress components
---from strain-squared, from vortex contraction, and from skew-strain---
contribute to the helicity flux. Note that the generalized Betchov relation
from
Appendix D can be applied to give
\be -  \langle {\rm Tr}\,\left(\ol{\bR}(\bS^{[k]})^2\right)\rangle =
   \frac{1}{2} \langle\ol{\bxi}^\top\bS^{[k]}\bomega^{[k]}\rangle
    + \frac{1}{4}\langle(\bomega^{[k]})^\top\ol{\bR}\bomega^{[k]}\rangle,
\lb{Betchov-3D-gen-H} \ee
where $\langle\cdot\rangle$ denotes a homogeneous average. Thus, the
strain-squared
contribution can be replaced on average with the above two terms.

It is known that the helicity cascade is local in scale (\cite{Eyink05}).
Therefore,
it is interesting to consider the $n=0$ contribution, which, assuming
$G=\varGamma,$
coincides with the helicity flux for the Nonlinear Model of the stress. Now,
it is not hard to see that
\be \ol{\bomega}^\top\ol{\bR}\ol{\bomega}=
\grad\bdot\left[(1/2)|\ol{\bomega}|^2\ol{\bomega}\right], \lb{om-R-om} \ee
which is a total derivative. Thus,
\be (1/4)\langle\ol{\bomega}^\top\ol{\bR}\ol{\bomega}\rangle=0,
   \lb{zero-om-R-om} \ee
and the contractile stress along vortex lines gives no contribution to
the UV-local part of mean helicity flux. Combining (\ref{zero-om-R-om})
and the generalized Betchov relation (\ref{Betchov-3D-gen-H}) gives also
\be -  \langle{\rm Tr}\,\left(\ol{\bR}(\ol{\bS})^2\right)\rangle =
   \frac{1}{2} \langle\ol{\bxi}^\top\ol{\bS}\ol{\bomega}\rangle.
\lb{Betchov-3D-gen-H-UV} \ee
Therefore, the total UV-local ($n=0$) contribution to mean helicity flux is
\be \langle\varLambda^{(0,1)}\rangle=  \frac{2}{3}C_2\ell^2\left\{
                      \frac{1}{2}
\langle\ol{\bxi}^\top\ol{\bS}\ol{\bomega}\rangle +
\langle\ol{\bR}\bdots(\ol{\bS}\btimes\ol{\bomega})\rangle\right\}
\lb{Hflux-NL3D} \ee
Equivalently, this is the Nonlinear Model expression for mean helicity flux.
The first term arises from the stress proportional to strain-squared and the
second term from the stress proportional to skew-strain.

The two terms in (\ref{Hflux-NL3D}) can be related by the following identity
\be \frac{1}{2} \ol{\bxi}^\top\ol{\bS}\ol{\bomega}=
     -\ol{\varbXi}\bdots(\ol{\bS}\btimes\ol{\bomega}). \lb{Xi-S-om}
\ee
This is easily proved by substituting on the right $\ol{\varXi}_{ij}=
-(1/2)\epsilon_{ijk}\ol{\xi}_k,$ then using the definition of the skew-strain
and the identity $\epsilon_{mij}\epsilon_{mkl}=\delta_{ik}\delta_{jl}
-\delta_{il}\delta_{jk}.$ Since
$(\grad\ol{\bomega})^\top=\ol{\bR}-\ol{\varbXi},$
using (\ref{Xi-S-om}) in (\ref{Hflux-NL3D}) gives
\be \langle\varLambda^{(0,1)}\rangle=  \frac{2}{3}C_2\ell^2
  \langle(\grad\ol{\bomega})^\top\bdots(\ol{\bS}\btimes\ol{\bomega})\rangle
\lb{Hflux-NL3D-II} \ee
Thus we see that both the symmetric and anti-symmetric parts of the
vorticity-gradient can contribute to helicity flux. The result
(\ref{Hflux-NL3D-II}) also makes clear the important role of the skew-strain
in the 3D helicity cascade. It is noteworthy that skew-strain makes no
UV-local contribution to energy flux at all and is thus free to adjust as
necessary to maintain the helicity flux in a joint cascade of both invariants.

If we use the result $(1/2)\ol{\bxi}^\top \bS^{[k]}\bomega^{[k]}=
-\ol{\varbXi}\bdots(\bS^{[k]}\btimes\bomega^{[k]})$ analogous to
(\ref{Xi-S-om}),
the generalized Betchov relation (\ref{Betchov-3D-gen-H}), and the expression
(\ref{Hflux-MM-3D1st}), then we obtain
\be \langle \varLambda^{(n,1)}_* \rangle
    = \frac{2}{3} \sum_{k=0}^n  \ol{C}_2^{[k]}\ell^2_k\left\{
    \langle
(\grad\ol{\bomega})^\top\bdots(\bS^{[k]}\btimes\bomega^{[k]})\rangle
 + \frac{1}{2} \langle(\bomega^{[k]})^\top\ol{\bR}(\bomega^{[k]})\rangle
\right\},
\lb{mean-Hflux-MM-3D1st} \ee
for the CSA expansion of mean helicity flux. This is analogous to the
similar result (\ref{mean-flux-MM-3D1st}) for the mean energy flux. However,
note that it is now the first term which makes a UV-local contribution
while the second only contributes for intermediate values of $k.$

We now consider a simple example to illustrate the mechanism of helicity
cascade by these stress terms:

\begin{Ex}
Vortex Tube Twisted by a Constant Screw
\end{Ex}

\begin{figure}
% \vspace{150pt}
\begin{center}
\includegraphics[width=250pt,height=200pt]{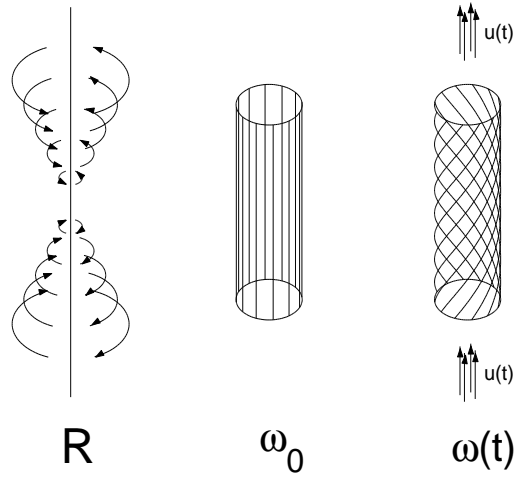}
\end{center}
\caption{{\small {\it Helicity Cascade by Vortex-Twisting}. The figure
illustrates
a cylindrical tube of parallel vortex lines (center) in a constant screw field
with twisting direction along the vortex axis (left). The result, shown on the
right, is that the vortex lines are twisted into helices. The tilting of the
small-scale vorticity vector and the solenoidal generation of an axial velocity
create helicity at small-scales. A positive (tensile) stress occurs, in the
plane perpendicular to the vortex axis at loose winding and along the twist
axis at tight winding.}}
\label{twist}
\end{figure}

We take as our model of the small-scales an exact stationary
solution of 3D Euler equations which was previously considered
by \cite{Mof} as an example of a helical flow with continuous
vorticity distribution. It is a two-dimensional but three-component
(2D-3C) velocity field, closely related to our previous Example 1.
Indeed, the horizontal components $(u^{[k]},v^{[k]})$ of the velocity
field are the same as those in (\ref{vortex-tube-vel}), obtained
from the 2D stream-function $\psi^{[k]}$ in (\ref{vortex-tube}).
However, this is now supplemented with a vertical velocity component
\be w^{[k]}(x,y)     = \left\{\begin{array}{cc}
                              (1/4)\omega_0 p[R^2-r^2] & r<R \cr
                                0 & r>R
                             \end{array} \right.  \lb{vortex-tube-w}  \ee
It can easily be shown that the resulting total velocity field is
a stationary Euler solution (e.g. see \cite{Mof}, Section 6(a).)
If $p=0,$ then this solution coincides with that in our Example 1.
The meaning of the parameter $p$ can best be understood by considering
the associated vorticity vector $\bomega^{[k]}=(\alpha^{[k]},\beta^{[k]},
\gamma^{[k]}).$ Of course, the vertical component $\gamma^{[k]}$ is
the same as in Example 1, $=\omega_0$ for $r<R$ and $=0$ for $r>R.$
The horizontal components are obtained using the vertical velocity
as a `stream function':
\be \left[\begin{array}{c}
          \alpha^{[k]} \cr
          \beta^{[k]}
          \end{array} \right]=\left[\begin{array}{c}
                                    \partial w^{[k]}/\partial y \cr
                                    -\partial w^{[k]}/\partial x \cr
                                    \end{array} \right]
       = \frac{1}{2}\omega_0 p\left[\begin{array}{c}
                                     -y \cr
                                      x
                                    \end{array} \right]
\lb{vortex-tube-alphabet} \ee
for $r<R$ and $=0$ for $r>R.$ The vortex lines are helices winding
around the $z$-axis with `pitch' $4\pi/p,$ i.e. making one counterclockwise
revolution in that vertical distance. It is not hard to check that the
solution given by (\ref{vortex-tube}),(\ref{vortex-tube-vel}),
(\ref{vortex-tube-w}), (\ref{vortex-tube-alphabet}) has a constant
helicity density $h^{[k]}=\bu^{[k]}\bdot\bomega^{[k]}=(1/4)\omega_0^2
R^2p$ for $r<R$ and $=0$ for $r>R.$

We shall take as our model of the large-scales the velocity
\be \ol{\bu}=\left[\begin{array}{c}
                   -\ol{\rho}yz/2 \cr
                   \,\,\,\,\ol{\rho}xz/2 \cr
                      0
                    \end{array} \right]. \ee
This corresponds to a constant {\it screw}, i.e. solid-body rotation
in each plane parallel to the $xy$-plane with an angular velocity
$\ol{\rho}z/2$ that grows linearly in $z.$ It is a righthand
screw for $\ol{\rho}>0$ and a lefthand screw for $\ol{\rho}<0.$
This velocity field has both non-vanishing strain and vorticity:
\be \ol{\bS}=\left[\begin{array}{ccc}
                   0 & 0 & -\ol{\rho}y/4 \cr
                   0 & 0 & \ol{\rho}x/4 \cr
                   -\ol{\rho}y/4 & \ol{\rho}x/4 & 0
                    \end{array} \right], \,\,\,\,
    \ol{\bomega}=\left[\begin{array}{c}
                   -\ol{\rho}x/2 \cr
                   -\ol{\rho}y/2 \cr
                    \ol{\rho}z \cr
                    \end{array} \right]. \lb{largeS-omega-screw} \ee
Furthermore, the vorticity-gradient matrix is constant and symmetric:
\be \grad\ol{\bomega}
             =\left[\begin{array}{ccc}
                   -\ol{\rho}/2 & 0 & 0 \cr
                   0 & -\ol{\rho}/2 & 0 \cr
                   0 & 0 & \ol{\rho}
                    \end{array} \right]=\ol{\bR}. \ee
It is not hard to check that the small-scale velocity field
defined previously becomes an exact time-dependent solution
of the `rapid-distortion equation' $(\partial_t+\bu\bdot\grad)
\bomega^{[k]}(t)=(\bomega^{[k]}(t)\bdot\grad)\bu=0,$ with $\bu=
\ol{\bu}+\bu^{[k]}(t),$ if the parameter $p=\ol{\rho}t.$ (This
is no longer equivalent to the full 3D Euler equation, since
the large-scales have also vorticity, neglected here.)
The small scales begin as the undisturbed vortex tube of
Example 1, whose filaments are then twisted into helices by the
large-scale screw. The resulting vorticity field of coiled helices
generates the axial velocity $w^{[k]}$ of (\ref{vortex-tube-w})
by solenoidal action, producing a net helicity in the small-scales.
See Figure 3.

Let us now consider the helicity transfer process, based upon
our general formulas for the stress. We shall only consider the
space region $r<R,$ since the small-scale fields outside the tube
are the same as for Example 1. Inside the small-scale vortex tube
the velocity-gradient tensor is
\be \bD^{[k]}(x,y) =   \frac{1}{2}\omega^{[k]}
                               \left[\begin{array}{ccc}
                                      0 & -1 & 0 \cr
                                      1 &  0 & 0 \cr
                                      -px &  -py & 0
                                     \end{array} \right]
\lb{twist-vortex-tube-D}  \ee
Substituting into (\ref{MNL-stress-diag-1st}) gives the stress
\be   \btau^{[k],(1)}_*     = \frac{1}{12}\ol{C}^{[k]}|\omega^{[k]}\ell_k|^2
                               \left[\begin{array}{ccc}
                                      1 &  0 & py \cr
                                      0 &  1 & -px \cr
                                      py  &  -px  & p^2r^2
                                     \end{array} \right]
\lb{twist-vortex-tube-rho}
\ee
to first order in gradients. Unlike Example 1, all three terms in
the stress formula (\ref{stress-NL3D}) [or (\ref{stress-NL3D-mat-dcmp})]
are present, including that from `skew-strain'. When $p$ is small
($p\ell_k\ll 1$) then this is essentially the same result as in Example 1,
but when $p$ is large ($p\ell_k\gg 1$) the dominant stress is tensile
along the screw axis. The resulting helicity flux has the form
\be   \Lambda^{[k],(1)}_*     =
      \frac{1}{6}\ol{C}^{[k]}\ol{\rho}|\omega^{[k]}\ell_k|^2
      \left[1-p^2r^2\right]
\lb{vortex-tube-Hflux}  \ee
When $p$ is small, there is a net transfer of helicity to the small-scales
of the same sign as the large-scale screw. This arises from the weakly
local transfer produced by the large-scale vorticity-gradient acting against
the contractile stress along the small-scale vortex. However, for large $p$ the
sign of helicity flux reverses, as the more tightly wound vortex lines produce
a net tensile stress along the screw axis by solenoidal action. The
contributions
to helicity flux of the separate
$\bS^{(n)}\bS^{(n)},[\varbOmega^{(n)},\bS^{(n)}],$
and $-\varbOmega^{(n)}\varbOmega^{(n)}$ stress terms in
(\ref{stress-NL3D-mat-dcmp})
are
\be  -\frac{1}{48}\ol{C}^{[k]}\ol{\rho}|p\omega^{[k]}\ell_k|^2r^2,\,\,\,\,
     -\frac{1}{8}\ol{C}^{[k]}\ol{\rho}|p\omega^{[k]}\ell_k|^2r^2,\,\,\,\,
     \frac{1}{6}\ol{C}^{[k]}\ol{\rho}|\omega^{[k]}\ell_k|^2
      \left[1-\frac{1}{8}p^2r^2\right],\,\,\,\,
\lb{Hflux-3}  \ee
respectively. We see that the flux comes mainly from the $-\varbOmega^{(n)}
\varbOmega^{(n)}$ term for $p\ell_k\ll 1$, as claimed above. For $p\ell_k\gg 1$
all three terms contribute, with the flux contribution from the `skew-strain'
six times bigger than that of either other term.

This example also illustrates the cascade of energy. Indeed, as was
observed already by \cite{BFLLM}, the transfer of helicity necessarily
involves also the transfer of energy. The formula (\ref{twist-vortex-tube-rho})
for the small-scale stress and (\ref{largeS-omega-screw}) for the large-scale
strain yield the following result for energy flux:
\be   \Pi^{[k],(1)}_*     =
      \frac{1}{24}\ol{C}^{[k]}\ol{\rho}p|\omega^{[k]}\ell_k|^2 r^2
\lb{twist-vortex-tube-Eflux}  \ee
There is a net forward transfer of energy if the signs of $\ol{\rho}$
and $p$ are the same. In that case, the large-scale screw winds the
small-scale helical vortex-lines more tightly, and kinetic energy is
generated in the axial velocity component by the resulting solenoidal
action. The contributions to energy flux of the separate $\bS^{(n)}\bS^{(n)},
[\varbOmega^{(n)},\bS^{(n)}],$ and $-\varbOmega^{(n)}\varbOmega^{(n)}$
stress terms in (\ref{stress-NL3D-mat-dcmp}) are
\be  0, \,\,\,\,\,\,\,\,
     \frac{1}{48}\ol{C}^{[k]}\ol{\rho}p|\omega^{[k]}\ell_k|^2 r^2,
     \,\,\,\,
     \frac{1}{48}\ol{C}^{[k]}\ol{\rho}p|\omega^{[k]}\ell_k|^2 r^2,
\lb{Eflux-3}  \ee
respectively. Inside the vortex, there is no energy transfer
from the strain-squared and instead equal amounts of the energy
flux are due to the contractile stress along vortex lines
and the stress proportional to `skew-strain'.

% \newpage

\section{Conclusions}

In this paper we have developed a novel approximation for turbulent
stress, via a multi-scale gradient (MSG) expansion. This scheme represents
the stress by an expansion in scales of motion and in orders of
space-gradients.
A major result (Appendix A) is that this expansion converges and furthermore
at a rapid rate for the `strongly local' part of the stress from the resolved
scales and adjacent subscales. However, the convergence of the spatial Taylor
expansion is expected to be much slower for stress contributions from scales
further below the filtering scale. Therefore, we have developed a more
approximate expansion, which should give a reasonable result at all scales with
just a few low-order velocity-gradients. This `coherent-subregions
approximation'
(CSA) is based on the assumption that the velocity increments across all the
subscale separation vectors should give a similar result, at a given scale,
as those for separation vectors in a `shell' where the gradient-expansion is
accurate and rapidly convergent.

An important application of our methods has been presented to the
three-dimensional
turbulent cascades of energy and helicity. Our main results are the formulas
(\ref{stress-NL3D}) for the stress, (\ref{flux-3D1st}) for the energy flux,
and (\ref{Hflux-3D1st}) for the helicity flux, exact to first-order in
gradients.
We have also developed the corresponding CSA expressions,
(\ref{MNL-stress-1st}),
(\ref{flux-MM-3D1st}), and (\ref{Hflux-MM-3D1st}), which are more heuristic but
which
should give a better representation of the very small-scale contributions than
the exact first-order results. We have generalized Betchov's well-known
relation,
which relates mean vortex-stretching and strain-skewness at the same scale,
to a similar relation between different scales ((\ref{Betchov-3D-gen}) and
Appendix D). This relation allowed us to derive expression
(\ref{mean-flux-MM-3D1st})
for mean energy flux and (\ref{mean-Hflux-MM-3D1st}) for mean helicity flux,
and
to analyze the expected contributions at different scales. Finally, we have
discussed
the physical mechanisms of energy and helicity cascade, in terms of our
analytical
formulas. We have shown by means of simple exact solutions of 3D Euler
equations
that our results are consistent with energy transfer by Taylor's mechanism of
`vortex-stretching' and with helicity transfer by a mechanism of
`vortex-twisting'.

There are many implications of the present work for experiment and simulation,
for theory, and for modelling.

A host of testable predictions have been provided for laboratory experiment
and for numerical simulation by our detailed formulas for turbulent stress,
energy flux, and helicity flux. The expansions in scale and in space have
been proved to converge, but the rate of convergence could be even faster
than what has been rigorously established and empirical studies can
determine this. The very distant subscales have been proved in
\cite{Eyink05} to give decreasing contributions to stress, and analytical
closures make further quantitative predictions about the mean amount
of energy and helicity transfer from each scale of motion (e.g. see
\cite{Kraichnan71b,AndreLesieur77}). Our multiscale formalism provides
a convenient framework within which to check these predictions, particularly
for experimentalists who cannot easily calculate spectral transfers. As to
the gradient-expansion, our first-order expressions (\ref{stress-NL3D}),
(\ref{flux-3D1st}), and (\ref{Hflux-3D1st}) should give good results
for the strongly local contributions, without further approximations.
Experiment and simulation can also check the validity of our CSA expansion,
which is based upon a bolder assumption. Assuming that our results are
empirically confirmed, experiment and simulation can also determine
the relative magnitudes of the various terms in our formulas. This
will help to shed further light on the detailed physical mechanisms
which underlie the turbulent cascades.

Our results suggest several further fruitful directions for theory.
Previous dynamical models of velocity-gradients (\cite{Vieillefosse82,
Vieillefosse84,Cantwell92,Chertkovetal99}) have investigated only
alignments between objects at the same scale. However, as our results
should make evident, alignments between velocity-gradients at distinct
scales are also of great importance in supporting turbulent cascades
and these inter-scale relations have received scant attention so far.
Improvement of our various approximations is another important avenue
for theory, particularly the CSA scheme, where many possible refinements
are apparent. Finally, theoretical methods to estimate subscale velocity
gradients from the resolved ones are strongly motivated by our work.
This would lead to LES modeling schemes, similar to those reviewed
in \cite{DomaradzkiAdams02}, but based upon a clearer picture of the
physical processes involved. If there is a universal mechanism which
generates small-scales from large-scales, then modeling this generation
process should lead to the most robust and generally applicable LES models.

A virtue of our approach is its wide range of potential applications.
Since it is based upon a very general feature of turbulent cascades---
i.e. their locality in scale and in space---the same scheme of
approximation can be exploited in many different situations, with,
of course, differing results depending upon the particular circumstances.
For example, in a following work (\cite{Ey05b}) we apply our methods
to the cascade of energy in 2D and obtain results consistent with
an inverse cascade. In this case, it is a weakly local interaction
via the `skew-strain' which plays the fundamental dynamical role.
We anticipate many other useful applications, such as passive scalar
cascades and magnetohydrodynamic cascades of energy and magnetic helicity.
We hope that our method will be useful for all these cases in illuminating
the physical mechanisms involved.

\vspace{1in}

\noindent {\bf Acknowledgements:} I wish to thank S. Chen, B. Ecke,
M. K. Rivera, M.-P. Wang, and Z. Xiao for a very fruitful collaboration
on 2D turbulence which helped to stimulate the development of the
general expansion method presented here. I would also like to thank
D. Carati, J. Domaradzki, A. Leonard, C. Meneveau and E. Vishniac
for very useful discussions and an anonymous referee for many
suggestions which helped to improve the paper. This work was
supported in part by NSF grant \# ASE-0428325.

\newpage

\appendix

\section{Convergence of the Gradient Expansion}

We establish here the convergence of the stress approximation $\btau^{(n,m)}$
in the limit $m\rightarrow\infty,$ for the space $L^1$-norm. The advantage
of this norm is that it allows us to give the proof by entirely elementary
methods (although convergence can doubtless be established as well using other
$L^p$ norms for $p>1$).
% We also give the proof, again for simplicity, in the
% unit periodic box ${\Bbb T}^d$ in $d$-dimensions.
We also give the proof, again for simplicity, in infinite volume in
$d$-dimensions
without flow boundaries.

Our argument uses the formula
\be \btau^{(n,m)} = \int d^d\br \,G_\ell(\br)
                \delta\bu^{(n,m)}(\br)\delta\bu^{(n,m)}(\br)
              - \int d^d\br \,G_\ell(\br) \delta\bu^{(n,m)}(\br)
                \int d^d\br \,G_\ell(\br) \delta\bu^{(n,m)}(\br),
\lb{tau-nm} \ee
with
\be \delta{\bu}^{(n,m)}(\br;\bx)
            = \sum_{p=1}^m \frac{1}{p!} (\br\bdot\grad)^p\bu^{(n)}(\bx).
\lb{bu-nm} \ee
If $\varGamma$ has a compactly-supported Fourier transform, then
$\bu^{(n)}(\bx)$
is real-analytic and thus $\delta{\bu}^{(n,m)}(\br;\bx)\rightarrow
\delta{\bu}^{(n)}(\br;\bx)$ as $m\rightarrow\infty$, pointwise in $\bx$
and also, as we see below, in the $L^1$ norm. Therefore, it is enough to
establish absolute integrability and summability:
\be \int d^d\br \,G_\ell(\br) \sum_{p,p'=0}^\infty \frac{1}{p!p'!}
    \|(\br\bdot\grad)^p\bu^{(n)}\,(\br\bdot\grad)^{p'}\bu^{(n)}\|_1<\infty
\lb{absolute-2} \ee
and
\be \int d^d\br \,G_\ell(\br)\int d^d\br' \,G_\ell(\br')
    \sum_{p,p'=0}^\infty \frac{1}{p!p'!}
    \|(\br\bdot\grad)^p\bu^{(n)}\,(\br'\bdot\grad)^{p'}\bu^{(n)}\|_1<\infty
\lb{absolute-1} \ee
In that case, the integrations and infinite summations commute and
%
%\begin{eqnarray}
%\lim_{m\rightarrow\infty}\btau^{(n,m)} & = & \int d^d\br \,G_\ell(\br)
%                \delta\bu^{(n)}(\br)\delta\bu^{(n)}(\br)
%              - \int d^d\br \,G_\ell(\br) \delta\bu^{(n)}(\br)
%                \int d^d\br \,G_\ell(\br) \delta\bu^{(n)}(\br) \cr
%        & = & \btau^{(n)}.
%\lb{commute}
%\end{eqnarray}
\begin{eqnarray}
\lim_{m\rightarrow\infty}\btau^{(n,m)} &=&  \int d^d\br \,G_\ell(\br)
                \delta\bu^{(n)}(\br)\delta\bu^{(n)}(\br)
              - \int d^d\br \,G_\ell(\br) \delta\bu^{(n)}(\br)
                \int d^d\br \,G_\ell(\br) \delta\bu^{(n)}(\br) \cr
     & = & \btau^{(n)}.
\lb{commute}
\end{eqnarray}

Let us establish the bound (\ref{absolute-2}). By the Cauchy-Schwartz
inequality,
\be \|(\br\bdot\grad)^p\bu^{(n)}\,(\br\bdot\grad)^{p'}\bu^{(n)}\|_1
    \leq |\br|^{p+p'} \|\grad^p\bu^{(n)}\|_2 \|\grad^{p'}\bu^{(n)}\|_2.
\ee
Thus, (\ref{absolute-2}) is implied by
\be \int d^d\br \,G_\ell(\br) \left[\sum_{p=0}^\infty \frac{1}{p!}
     |\br|^p \|\grad^p\bu^{(n)}\|_2\right]^2 <\infty
\lb{absolute-sq} \ee
By a similar argument, we see that (\ref{absolute-1}) holds, if
\be \int d^d\br \,G_\ell(\br) \left[\sum_{p=0}^\infty \frac{1}{p!}
     |\br|^p \|\grad^p\bu^{(n)}\|_2\right] <\infty
\lb{absolute-un} \ee
%
% To proceed we must have an estimate of $\|\grad^p\bu^{(n)}\|_2$. This is
% easy to obtain by going over to Fourier coefficients using the Parseval
% equality:
% \be \|\grad^p\bu^{(n)}\|_2^2 = \sum_{\bk} \left| (i\bk)^p \,
%       \widehat{\varGamma}(\ell_n\bk)\,\widehat{\bu}(\bk)\right|^2.
% \lb{Parseval} \ee
% Since $\widehat{\varGamma}$ has compact support, the summation over $\bk$
%%involves
% only wavenumbers satisfying $|\bk|<c_1/\ell_n$ for some constant $c_1.$
%%Thus,
% \be \|\grad^p\bu^{(n)}\|_2^2 \leq c_2^2 (c_1/\ell_n)^{2p} \sum_{\bk}
%                             |\widehat{\bu}(\bk)|^2,
% \lb{gradu-ineq-1} \ee
To proceed we must have an estimate of $\|\grad^p\bu^{(n)}\|_2$. This is
easy to obtain by going over to Fourier transforms using the Plancherel
identity:
\be \|\grad^p\bu^{(n)}\|_2^2 = \int d^d\bk \left| (i\bk)^p \,
       \widehat{\varGamma}(\ell_n\bk)\,\widehat{\bu}(\bk)\right|^2.
\lb{Parseval} \ee
Since $\widehat{\varGamma}$ has compact support, the integral over $\bk$
involves
only the wavenumbers with $|\bk|<c_1/\ell_n$ for some constant $c_1.$  Thus,
\be \|\grad^p\bu^{(n)}\|_2^2 \leq c_2^2 (c_1/\ell_n)^{2p} \int d^d\bk
                             \,|\widehat{\bu}(\bk)|^2,
\lb{gradu-ineq-1} \ee
for another constant $c_2=\sup_\bk |\widehat{\varGamma}(\bk)|,$ or
\be \|\grad^p\bu^{(n)}\|_2  \leq c_2 (c_1/\ell_n)^{p} \|\bu\|_2.
\lb{gradu-ineq-2} \ee
With this estimate we get
\be \sum_{p=0}^\infty \frac{1}{p!} |\br|^p \|\grad^p\bu^{(n)}\|_2 \leq
                            c_2 \exp(c_1 |\br|/\ell_n) \|\bu\|_2.
\lb{summation-ineq} \ee
Thus, we see that (\ref{absolute-sq}) and (\ref{absolute-un}) hold, if the
filter kernel $G(\br)$ decays faster than exponentially in space. For example,
$G$ could be Gaussian. Notice that $\varGamma(\br)$ can also decay very rapidly
in space---for example, faster than any inverse power---since the Fourier
transform $\widehat{\varGamma}(\bk)$ may be both compactly supported and
$C^\infty.$
For any such filter kernels $G$ and $\varGamma$, we conclude finally that the
absolute summability/integrability conditions
(\ref{absolute-2}),(\ref{absolute-1})
both hold, and thus the limit relation (\ref{commute}) is valid, as claimed.
QED.

\section{Comparison with the Defiltering Approach}

As mentioned in the Introduction, \cite{YeoBedford88}, \cite{Leonard97}, and
\cite{Caratietal01} have developed a somewhat similar gradient expansion
for the turbulent stress. Here we would like to compare and contrast the two
approaches and, in particular, indicate our reasons for dissatisfaction with
the expansion constructed by those authors. Their approach is based upon
{\it defiltering}, which is the inverse to the filtering operator,
defined spectrally by
\be {\cal G}_\ell \bv(\bx)  = \int d^d\bk \,\, \widehat{G}(\bk\ell)
\widehat{\bv}(\bk)
                                 e^{i\bk\bdot \bx}. \lb{filter-op} \ee
Thus, the defiltering operator is given similarly by
\be {\cal G}^{-1}_\ell\bv(\bx)  = \int d^d\bk \,\, [\widehat{G}(\bk\ell)]^{-1}
                     \widehat{\bv}(\bk) e^{i\bk\bdot \bx}. \lb{defilter-op} \ee
In terms of these operators, \cite{YeoBedford88,Leonard97} and
\cite{Caratietal01}
define a tensor-valued functional
\be \bT[\bv] \equiv
        {\cal G}_\ell\{{\cal G}^{-1}_\ell\bv \,\,{\cal G}^{-1}_\ell\bv \}
       -\bv \,\,\bv. \lb{T-fun} \ee
A little thought shows that if $\ol{\bu}={\cal G}_\ell\bu$ is substituted into
this functional, then one recovers the turbulent stress as
$\btau=\bT[\ol{\bu}].$
Hence, this formula provides, seemingly, an exact closure of the stress in
terms
of the filtered velocity $\ol{\bu}.$ Furthermore, the functional $\bT$ has a
formal
gradient expansion:
\be \sT_{ij}[\bv] = \sum_{{\bmi p},{\bmi q}} c_{{\bmi p},{\bmi q}} \grad^{\bmi
p}v_i
                     \,\,\grad^{\bmi q}v_j, \lb{T-exp} \ee
where the summation is over multi-indices ${\bmi p}=(p_1,...,p_d),{\bmi
q}=(q_1,...,q_d)$
with integer components. Note that $\grad^{\bmi p}=\partial_{1}^{p_1}\cdots
\partial_{d}^{p_d}$
for the multi-index ${\bmi p}.$ The coefficients in the expansion (\ref{T-exp})
can be
obtained from a generating function (\cite{Caratietal01}):
$$  F[\bphi,\bpsi]\equiv \frac{\widehat{G}(-i(\bphi+\bpsi))}
                   {\widehat{G}(-i\bphi)\widehat{G}(-i\bpsi)}
                  = \sum_{{\bmi p},{\bmi q}} c_{{\bmi p},{\bmi q}}
                    \bphi^{\bmi p} \bpsi^{\bmi q}.  $$
Here $\bphi^{\bmi p}=\phi_{1}^{p_1}\cdots \phi_{d}^{p_d}$. Thus, the expansion
(\ref{T-exp}) for $\btau=\bT[\ol{\bu}]$ seems to yield a closed, constitutive
formula for the turbulent stress in terms of the gradients of the filtered
velocity.

To see the problem with this approach, note that the defiltering operator
is not even defined if the filter kernel has a compactly supported Fourier
transform.
In this case, all the subscale modes cannot be recovered from knowledge of the
filtered velocity $\ol{\bu}$. However, this is only an extreme form of a
general
difficulty. For any filter kernel $G,$ the defiltering operator ${\cal
G}_\ell^{-1}$
is {\it unbounded} on the natural function spaces for the velocity field, such
as the $L^p$ spaces. This means that defiltering is not defined for every
element
of those spaces, but instead only for a (dense) subspace, called the `domain'
of the operator. The natural, maximal domain of the defiltering operator in any
of these
spaces is the range of the corresponding filtering operator on that space, or
${\rm Dom}
({\cal G}_\ell^{-1})={\rm Ran}({\cal G}_\ell).$ This means that, to defilter a
function
in the space, that element must have been obtained by filtering another member
of
the space. However, this is not the case for most functions in the space, even
infinitely smooth ones.

For example, consider the most natural space of $L^2$ or finite-energy fields,
and
a Gaussian filter kernel $G.$ Most $\bv\in L^2$, even those with very rapidly
decaying
Fourier coefficients $\widehat{\bv}(\bk)$, have no Gaussian defilter. Formally,
\be [{\cal G}_\ell^{-1}\bv]\,\widehat{\,\,}\,(\bk)
                    = \exp[+(k\ell)^2/2\sigma^2]\widehat{\bv}(\bk)
\lb{Gauss-defilter} \ee
for a Gaussian kernel with Fourier transform
$\widehat{G}(\bk)=\exp(-k^2/2\sigma^2).$
Thus, the Fourier coefficients $\widehat{\bv}(\bk)$ might decay very rapidly,
e.g. exponentially in $|\bk|,$ and yet the defiltered field has infinite energy
or ${\cal G}_\ell^{-1}\bv\notin L^2.$ For such a velocity field, the stress
$\bT[\bv]$ defined by the formula (\ref{T-fun}) does not exist. This poses
a real difficulty for an LES closure equation based upon a constitutive
relation $\btau=\bT^{(m)}[\ol{\bu}]$ obtained by truncating the expansion
(\ref{T-exp}) at finite order $m$. There is nothing to guarantee
that the solution $\ol{\bu}$ of such a closure equation will have Fourier
coefficients decaying fast enough for (\ref{Gauss-defilter}) to remain in
$L^2$.
In that case, the expansion (\ref{T-exp}) will not converge and there is no
reason
to expect that the solution of the $m$th-order LES equation will converge
in the limit $m\rightarrow\infty$ to the exact filtered velocity, even though
the closure then becomes formally `exact'. Of course, even worse,
the LES equation may itself be ill-posed and its solution could blow up at
finite
time. The `exactness' of the closure as $m\rightarrow\infty$ does not provide
any guarantee of good behavior at finite $m.$

Such difficulties with defiltering are not unknown in the LES literature.
\cite{DomaradzkiAdams02} have reviewed various approaches to subgrid stress
modelling by defiltering and have pointed out the related fact that defiltering
is generally an ill-posed operation in function space. In particular,
multiplication
by the inverse filter transform, as in our equation (\ref{Gauss-defilter}),
magnifies
the effects of noise and round-off error at high-wavenumbers. Thus, the
defiltering
operation, even when it exists, is not stable to small perturbations in the
input
velocity field.  As in any ill-posed problem, various regularizations may be
considered
to render it well-posed. Most of the existing approaches have employed an
approximate,
regularized defiltering together with an ``eddy-viscosity'' or dissipative term
at high-wavenumbers. See \cite{DomaradzkiAdams02}. Needless to say, our
approach
is quite different. We have constructed an approximation scheme which is proved
to converge under very modest assumptions but which is not closed. Our present
goal is to develop a tool to explore the basic physics and not to construct
turbulence models directly. Our expressions for turbulent stress may be useful
in modeling efforts, but we must defer to future work the important problem
of estimating the unknown subscale gradients that appear.

\section{CSA Expansion Coefficients for Gaussian Kernel}

We give here the coefficients that appear in the CSA expansion, for the
special case of an isotropic  Gaussian filter:
$$  G_\ell(\br) =\frac{1}{[2\upi(\sigma\ell)^2]^{d/2}}
\exp\left[-\frac{r^2}{2(\sigma\ell)^2}\right]. $$
In general, one would like to have $\sigma\approx 1,$ so that this really
corresponds
to a filter at scale $\ell$. However, this is often not true; for example, the
conventional
choice made by \cite{Leonard74} was $\sigma^2=1/12$ (so that the second moment
of the Gaussian and box filters would agree). In such cases it is better to
define the
`shells' in the model formulation as
\be {\mathcal S}_0 = \{\br:\,|\br|>\sigma\ell_0\} \lb{shell-0} \ee
\be {\mathcal S}_k =
\{\br:\,\sigma\ell_{k-1}>|\br|>\sigma\ell_k\},\,\,\,k=1,...,n
\lb{shell-k} \ee
In this way, increments for separations $r\approx \sigma\ell_n$ are calculated
from
fields $\bu^{(n)}$ filtered at the same length-scale. We must calculate the
partial
averages of $|\br|^p,\,\,p=2,4,6...$ with respect to the Gaussian filter, over
each
of these shells. Introducing a dimensionless variable $\brho=\br/\ell$, these
moment
averages may be written as $C_p^{[k]}\ell^p$ for shell ${\mathcal S}_k
,\,\,k=0,1,...,n.$

The integrals for coefficients $C_p^{[k]}$ are evaluated by substituting
$t=\rho^2/(2\sigma^2), \sigma^2 dt= \rho d\rho,$ and using
$S_{d-1}=2\upi^{d/2}/
\Gamma(d/2)$ for the $(d-1)$-volume of the unit sphere in dimension $d$:
\begin{eqnarray}
C_p^{[0]} & = & S_{d-1}\int_\sigma^\infty  \rho^{d+p-1}
                \frac{e^{-\rho^2/2\sigma^2}}{(2\upi\sigma^2)^{d/2}} \, d\rho
\cr
        \,& = & \frac{(2\sigma^2)^{p/2}}{\Gamma(d/2)}
                \int_{1/2}^\infty t^{(d+p)/2-1} e^{-t} \, dt
             =  \frac{2^{p/2}}{\Gamma(d/2)}
                 \Gamma\left(\frac{d+p}{2},\frac{1}{2}\right)\sigma^p,
\lb{coeff-0}
\end{eqnarray}
and
\begin{eqnarray}
C_p^{[k]} & = & S_{d-1}\int_{\sigma\lambda^{-k}}^{\sigma\lambda^{-(k-1)}}
\rho^{d+p-1}
                \frac{e^{-\rho^2/2\sigma^2}}{(2\upi\sigma^2)^{d/2}} \, d\rho
\cr
%        \,& = & \frac{S_{d-1}}{\upi^{d/2}} \sigma^p
%                \int_{1/2\lambda^{2n}}^{1/2\lambda^{2(n-1)}} t^{(d+p)/2-1}
%%e^{-t} \, dt \cr
        \,& = &  \frac{2^{p/2}}{\Gamma(d/2)} \left[
                 \gamma\left(\frac{d+p}{2},\frac{1}{2\lambda^{2(k-1)}}\right)
-\gamma\left(\frac{d+p}{2},\frac{1}{2\lambda^{2k}}\right)\right] \sigma^p,
\lb{reg-coeff} \end{eqnarray}
for $k=1,...,n$.
% \begin{eqnarray}
% C_p^{(N)} & = & S_{d-1}\int_{0}^{\sigma\lambda^{-(N-1)}} \rho^{d+p-1}
%                \frac{e^{-\rho^2/2\sigma^2}}{(2\upi\sigma^2)^{d/2}} \, d\rho
%%\cr
%        \,& = & \frac{S_{d-1}}{\upi^{d/2}} \sigma^p
%                \int_{1/2\lambda^{2n}}^{1/2\lambda^{2(n-1)}} t^{(d+p)/2-1}
%%e^{-t} \, dt \cr
%        \,& = &  \frac{S_{d-1}}{\upi^{d/2}} \left[
%                 \Gamma\left(\frac{d+p}{2}\right)-
%
%%\Gamma\left(\frac{d+p}{2},\frac{1}{2\lambda^{2(N-1)}}\right)\right] \sigma^p.
%         \,& = & \frac{2^{p/2}}{\Gamma(d/2)}
%
%%\gamma\left(\frac{d+p}{2},\frac{1}{2\lambda^{2(N-1)}}\right) \sigma^p
% \end{eqnarray}
Here we have introduced the incomplete Gamma functions $\Gamma(a,x)$ and
$\gamma(a,x)
=\Gamma(a)-\Gamma(a,x),$ as they are defined in the standard literature
(\cite{AbramSteg}).

The asymptotics of these coefficients for large $k$ are obtained using
$\gamma(a,x)\sim x^a/a$ as $x\rightarrow 0$ (e.g. \cite{AbramSteg}, 6.5.4 \&
6.5.29).
Thus,
$$ C_p^{[k]} \sim C(d,p,\lambda) \lambda^{-(d+p)k}
\sigma^p,\,\,\,\,k\rightarrow \infty, $$
with
$C(d,p,\lambda)=(\lambda^{d+p}-1)/\left[(d+p)\Gamma(d/2)2^{(d-2)/2}\right].$ If
we
define $\overline{C}_p^{[k]}= \lambda^{(d+p)k} C_p^{[k]},$ then this new
constant
becomes independent of $k$ as $k\rightarrow\infty$:
\be \overline{C}_p^{[k]}\sim C(d,p,\lambda)\sigma^p.\lb{Cbar-asymp} \ee
It is not hard to prove that these same asymptotics hold, at least as big-$O$
bounds,
for much more general filter kernels than Gaussian.

An interesting application of (\ref{Cbar-asymp}) is to establish the order of
magnitude of the `systematic' stress term $\bvrho^{[k],(m)}_{*}$ that appears
in
equation (\ref{MNL-stress-I}) for $\btau^{[k],(m)}_{*}.$ We assume that the
velocity field $\bu$ has the H\"{o}lder exponent $0<\alpha<1.$ From the
definition
(\ref{missing-1b}) it is obvious that $\bvrho^{[k],(m)}_{*}$ is a sum over
$p,p'=0,...,m$ of terms proportional to
\be \overline{C}_{p+p'}^{[k]}\ell^{p+p'}_k
(\partial^p\bu^{[k]})(\partial^{p'}\bu^{[k]}) \lb{Cpp'-term} \ee
Here $\partial^p\bu^{[k]}$ indicates a $p$th-order space-derivative of
$\bu^{[k]}$ with
component indices suppressed. It is not hard to show (e.g. see \cite{Eyink05})
that
$\partial^p\bu^{[k]}=O(\ell_k^{\alpha-p}).$ On the other hand,
(\ref{Cbar-asymp})
implies that $\overline{C}_{p+p'}^{[k]}$ is asymptotically constant as
$k\rightarrow\infty.$
It follows that each term $p,p'$ of the Taylor expansion contributes a term for
$\bvrho^{[k],(m)}_*$ that scales as $O(\ell_k^{2\alpha})$. In fact, this is the
correct
order of magnitude for the contribution to the stress from length-scale
$\ell_k$ (\cite{Eyink05}).
It is worth pointing out that the term
$\bu^{\prime\,[k],(m)}_*\bu^{\prime\,[k'],(m)}_*$
in (\ref{MNL-stress-I}) would likewise scale as
$O(\ell_k^{\alpha}\ell_{k'}^{\alpha}),$
{\it if} the correction factor $N_k$ had been used in the definition
(\ref{missing-2b})
rather than $N_k^{1/2}.$ Indeed, so defined,
$\bu^{\prime\,[k],(m)}_*\bu^{\prime\,[k'],(m)}_*$
is from (\ref{missing-1b}) a sum over $p,p'$ of terms proportional to
$\overline{C}_{p}^{[k]}\overline{C}_{p'}^{[k']}\ell^{p}_k\ell^{p'}_{k'}
(\partial^p\bu^{[k]})(\partial^{p'}\bu^{[k']}).$ In that case, 
the previous argument carries through.
It might seem that this alternative definition of $\bu^{\prime\,[k],(m)}_*$
therefore
has some merit, except that it ignores the cancellations that we expect to
occur
in the integrals over volume.

% Then,
% $$ C_2^{(0)} = 3\sigma^2\exp\left(-\frac{1}{2}\right), $$
% $$ C_2^{(n)} =
%%\sigma^2\left[\left(2+\frac{1}{2^{2n}}\right)
% \exp\left(-\frac{1}{2^{2n+1}}\right)
%             -
%%\left(2+\frac{1}{2^{2(n-1)}}\right)\exp\left(-\frac{1}{2^{2n-1}}
% \right)\right],
%   \,\,\,\,n=1,...,N-1$$
% $$ C_2^{(N)} = \sigma^2\left[2- \left(2+\frac{1}{2^{2(N-1)}}\right)
%
%%\exp\left(-\frac{1}{2^{2N-1}}\right)\right]. $$
% and
% $$ C_4^{(0)} = 13\sigma^4\exp\left(-\frac{1}{2}\right), $$
% $$ C_4^{(n)} =
%%\sigma^4\left[\left(8+\frac{1}{2^{2(n-1)}}+\frac{1}{2^{4n}}\right)
%                \exp\left(-\frac{1}{2^{2n+1}}\right)
%             - \left(8+\frac{1}{2^{2(n-2)}}+\frac{1}{2^{4(n-1)}}\right)
%                \exp\left(-\frac{1}{2^{2n-1}}\right)\right], $$
% $$
%%\,\,\,\,\,\,\,\,\,\,\,\,\,\,\,\,\,\,\,\,\,\,
% \,\,\,\,\,\,\,\,\,\,\,\,\,\,\,\,\,\,\,\,\,\,\,\,\,\,
%    \,\,\,\,\,\,\,\,\,\,\,\,\,\,\,\,\,\,\,\,\,\,\,\,\,\,\,\,n=1,...,N-1$$
% $$ C_4^{(N)} = \sigma^4\left[8 -
%%\left(8+\frac{1}{2^{2(N-2)}}+\frac{1}{2^{4(N-1)}}\right)
%                \exp\left(-\frac{1}{2^{2N-1}}\right)\right]. $$
% Notice that for large $n,$ the following asymptotics holds
% $$ C_2^{(n)}\sim \frac{15}{4} \sigma^2 2^{-4n},\,\,\,\, C_4^{(n)}\sim
%%\frac{21}{2} \sigma^4 2^{-6n}.$$

\section{Generalized Betchov Relation in 3D}

We here give briefly the proof of the generalized Betchov relation
(\ref{Betchov-3D-gen}).
Let us suppose that ${\bmi a},{\bmi b},{\bmi c}$ are three solenoidal
(divergence-free)
vector fields in 3D. Then, it is trivial to verify from the product rule of
differentiation that
\be a_{i,j}b_{j,k}c_{k,i} + c_{k,j}b_{j,i}a_{i,k} =
\partial_i(a_{i,j}b_{j,k}c_k)- \partial_k(a_{i,j}b_{j,i}c_k) +
\partial_j(a_{i,k}b_{j,i}c_k)
\lb{prod-rule} \ee
where we use the notation $a_{i,j}=\partial a_i/\partial x_j,$ etc. Because the
righthand
side is a total space-derivative, the ensemble-average of the lefthand side is
zero if one
assumes space-homogeneity:
\be \langle a_{i,j}b_{j,k}c_{k,i} + c_{k,j}b_{j,i}a_{i,k}\rangle = 0.
\lb{abc-Betch} \ee
This same relation holds for space-averages, if boundary conditions on ${\bmi
a},{\bmi b},{\bmi c}$
are such that boundary terms from integration by parts can be ignored. With
either of these
assumptions, let us then apply (\ref{abc-Betch}) for ${\bmi b}=\ol{\bu}$ and
${\bmi a}={\bmi c}
=\bu^{(n)}$. This gives
\be \langle {\rm Tr}[\ol{\bD}(\bD^{(n)})^2]\rangle =0, \lb{DDD-Betch} \ee
with $\sD_{ij}=u_{i,j}$ the deformation matrix associated to a velocity field
$\bu.$

Now, using $\sD_{ij}^{(n)}=\sS^{(n)}_{ij}-(1/2)\epsilon_{ijk}\omega^{(n)}$
[equation (\ref{gradu-S-om})] gives
\begin{eqnarray}
(\bD^{(n)})^2  & = & (\bS^{(n)})^2
       -\frac{1}{4}(\bI|\bomega^{(n)}|^2-\bomega^{(n)}\bomega^{(n)})\cr
                     &   & \,\,\,\,\,\,
  +\frac{1}{2}(\bomega^{(n)}\btimes\bS^{(n)}+\bS^{(n)}\btimes \bomega^{(n)}).
\lb{Dn-sq} \end{eqnarray}
Note that the matrices on the first line of the righthand side are symmetric
and
the matrix on the second line is antisymmetric. ($\bI$ is the identity matrix.)
Thus, if one uses the relation analogous to (\ref{gradu-S-om}) for
$\ol{\sD}_{ij},$
then the only contribution to the trace of $\ol{\bD}$ with $(\bD^{(n)})^2$ is
from the trace of $\ol{\sS}_{ij}$ with the first line in (\ref{Dn-sq}) and from
the
trace of $(1/2)\epsilon_{ijk}\ol{\omega}_k$ with the second line. Using
tracelessness of the strain matrix $\ol{\bS}$ and the identity $\epsilon_{ijk}
\epsilon_{ilm}=\delta_{jl}\delta_{km}-\delta_{jm}\delta_{kl}$ gives
\be {\rm Tr}[\ol{\bD}(\bD^{(n)})^2]= {\rm
Tr}\,\left(\ol{\bS}(\bS^{(n)})^2\right)
   + \frac{1}{4}(\bomega^{(n)})^\top\ol{\bS}(\bomega^{(n)})
   + \frac{1}{2}(\ol{\bomega})^\top\bS^{(n)}(\bomega^{(n)}). \lb{Tr-DDD} \ee
Substituting (\ref{Tr-DDD}) into (\ref{DDD-Betch}) gives
(\ref{Betchov-3D-gen}). QED.

% Proof of the `helical' Betchov relation:
%
% $$ \partial_i[\omega_j(\bS\btimes\bomega)_{ij}] =
%%\partial_i[\epsilon_{jkl}\omega_{j}\sS_{ik}\omega_l]
%   = \epsilon_{jkl}(\omega_{j,i}\sS_{ik}\omega_l + \omega_j\sS_{ki,i}\omega_l
%                    +\omega_j \sS_{ik}\omega_{l,i})$$
% Note $\epsilon_{jkl}\omega_j\omega_l \sS_{ki,i}=0.$ Thus,
% $$ \langle \epsilon_{jkl}\omega_{j,i}\sS_{ik}\omega_l\rangle =
%   -\langle \epsilon_{jkl}\omega_{j}\sS_{ik}\omega_{l,i}\rangle =
%   -\langle \epsilon_{lkj}\omega_{l}\sS_{ik}\omega_{j,i}\rangle =
%   \langle \epsilon_{jkl}\omega_{j,i}\sS_{ik}\omega_{l}\rangle
% $$
% Hence,
% $$ \langle \bR\bdots\,(\bS\btimes\bomega)\rangle =
%   \frac{1}{2}\langle
%%\epsilon_{jkl}(\omega_{i,j}+\omega_{j,i})\sS_{ik}\omega_l\rangle=  $$

\end{document}